\documentclass[manuscript,screen]{acmart}

\usepackage[utf8]{inputenc}
\usepackage[english]{babel}

\usepackage{graphicx}
\DeclareGraphicsExtensions{.pdf,.png,.jpg}

\usepackage{booktabs}
\usepackage{multirow}
\usepackage{enumitem}
\usepackage{makecell}
\usepackage{relsize}
\usepackage{caption}
\usepackage{tabularx}
\usepackage{array}
\usepackage{pifont}
\usepackage{longtable}
\usepackage{pdflscape}
\usepackage{xcolor}
\usepackage{ragged2e}
\usepackage{colortbl}
\usepackage{microtype}
\usepackage{xltabular}
\usepackage{adjustbox}
\usepackage{cprotect}
\usepackage{seqsplit}
\usepackage{fvextra}
\usepackage{needspace}
\usepackage[T1]{fontenc}
\usepackage{wrapfig}

\usepackage[most]{tcolorbox}
\usepackage[normalem]{ulem}

\definecolor{darkgreen}{RGB}{0,100,0}
\definecolor{codegreen}{rgb}{0,0.5,0}
\definecolor{codepurple}{rgb}{0.58,0,0.82}
\definecolor{codegray}{rgb}{0.5,0.5,0.5}
\usepackage{listings}
\lstdefinestyle{mystyle}{
  commentstyle=\color{codegreen},
  keywordstyle=\bfseries,
  stringstyle=\color{codepurple},
  basicstyle=\ttfamily\scriptsize,
  breaklines=true,
  captionpos=b,
  keepspaces=true,
  tabsize=2
}
\usepackage[ruled,vlined,linesnumbered]{algorithm2e}

\usepackage{enumitem}

\usepackage{tikz}
\usetikzlibrary{arrows.meta,positioning,shapes,calc} 

\usepackage{mdframed}

\definecolor{codebg}{gray}{0.97}
\definecolor{codecomment}{gray}{0.40}
\definecolor{codekw}{rgb}{0.0,0.0,0.55}
\lstdefinestyle{kernelC}{
  language=C,
  basicstyle=\ttfamily\small,
  commentstyle=\color{codecomment}\itshape,
  keywordstyle=\color{codekw}\bfseries,
  backgroundcolor=\color{codebg},
  numbers=none,
  xleftmargin=4pt,
  xrightmargin=4pt,
  frame=single,
  framerule=0.4pt,
  rulecolor=\color{black!30},
  breaklines=true,
  breakatwhitespace=false,
  columns=flexible,
  keepspaces=true,
  showstringspaces=false,
  tabsize=4,
  aboveskip=6pt,
  belowskip=6pt,
  morekeywords={SEC,bpf_ringbuf_reserve,bpf_ringbuf_submit,
    __xdp_return,skb_frag_page,skb_frag_size,page_address,
    xsk_buff_get_tail,xsk_buff_del_tail,update_effective_progs,
    list_del,bpf_link_free,purge_effective_progs,
    MEM_TYPE_XSK_BUFF_POOL,PTR_OR_NULL,bpf_xdp_adjust_tail,
    bpf_xdp_shrink_data},
  literate={→}{$\rightarrow$}1 {←}{$\leftarrow$}1
           {↑}{$\uparrow$}1 {↔}{$\leftrightarrow$}1,
}

\lstdefinestyle{diagram}{
  basicstyle=\ttfamily\small,
  backgroundcolor=\color{codebg},
  numbers=none,
  xleftmargin=4pt,
  xrightmargin=4pt,
  frame=single,
  framerule=0.4pt,
  rulecolor=\color{black!30},
  breaklines=true,
  columns=flexible,
  keepspaces=true,
  aboveskip=6pt,
  belowskip=6pt,
  literate={→}{$\rightarrow$}1 {←}{$\leftarrow$}1
           {↑}{$\uparrow$}1,
}

\lstdefinestyle{pythoncode}{
  style=diagram,
  language=Python,
  basicstyle=\ttfamily\scriptsize,
  keywordstyle=\color{blue!70!black}\bfseries,
  commentstyle=\color{green!40!black}\itshape,
  stringstyle=\color{red!60!black},
  showstringspaces=false
}

\newcommand{\find}[1]{%
\begin{tcolorbox}[tile,size=fbox,boxsep=2mm,boxrule=0pt,top=0pt,bottom=0pt,
borderline={0.6mm}{0pt}{black!66!white},colback=black!5!white]
\em #1
\end{tcolorbox}
}

\definecolor{revisionblue}{RGB}{80,140,210}

\newboolean{COMMENTSON} 
\setboolean{COMMENTSON}{true}   
\ifthenelse{\boolean{COMMENTSON}}
{

}

\definecolor{DarkOrange}{rgb}{0.8,0.3,0.0} 
\definecolor{DarkCyan}{rgb}{0.0, 0.55, 0.55}
\definecolor{codegreen}{rgb}{0,0.6,0}
\definecolor{codegray}{rgb}{0.5,0.5,0.5}
\definecolor{codepurple}{rgb}{0.58,0,0.82}
\definecolor{backcolour}{rgb}{0.95,0.95,0.92}

\newcommand{\papertitle}{Python Import as an Execution Boundary: An Empirical Study of Bugs, Vulnerabilities, and Analysis Gaps}
\newcommand{\papertitleshort}{Python Import as an Execution Boundary: An Empirical Study of Bugs, Vulnerabilities, and Analysis Gaps}

\newcommand{\tech}{\mbox{\textsc{ImportMine}}}

\newcommand{\bench}{\mbox{\textsc{ImportVulBench}}}

\newcommand{\paperkeywords}{Python, import security, static analysis, software supply chain, vulnerability study}

\newcommand{\authorAname}{Baihong Chen}
\newcommand{\authorAaffil}{Utah State University}
\newcommand{\authorAemail}{b.chen@usu.edu}

\newcommand{\authorBname}{Wen Li}
\newcommand{\authorBaffil}{Utah State University}
\newcommand{\authorBemail}{awen.li@usu.edu}

\newcommand{\circleone}[1]{%
  \tikz[baseline=(char.base)]{
    \node[
      shape=circle,
      draw=black,
      fill=white,
      text=black,
      inner sep=1pt
    ] (char) {\small #1};
  }%
}

\newcommand{\artifact}[1]{%
  \href{https://github.com/BaihongChen/ImportMine}{\underline{#1}}%
}

\setcopyright{none}
\renewcommand\footnotetextcopyrightpermission[1]{}

\begin{document}

\title[\papertitleshort]{\papertitle}

\author{\authorAname}
\affiliation{%
  \institution{\authorAaffil}
  \city{Logan}
  \state{Utah}
  \country{USA}
}
\email{\authorAemail}
\authornote{First author}

\author{\authorBname}
\affiliation{%
  \institution{\authorBaffil}
  \city{Logan}
  \state{Utah}
  \country{USA}
}
\email{\authorBemail}
\authornote{Corresponding author}


\begin{abstract}

Python import does more than resolve dependencies: it executes code during
module and package initialization. This behavior can trigger failures, load
dynamic or native code, access resources, or change security-sensitive state
before an application calls a package API. Prior work studies package
selection, malicious packages, or package vulnerabilities.

We present {\tech}, a study of import-related bugs and security vulnerabilities in
Python software. We combine security advisories with PyPI project
histories and use source and patch evidence to confirm how import activates
cases, why the problem occurs, how developers fix it, and what program
information is needed to explain the behavior. 
We retain 31 import-related advisory vulnerabilities and 38 application-data
boundary cases and confirm 1,429 project-history bugs across 1,302 repositories.
Among the project-history bugs activated during initialization, 97.6\%
stop or disrupt normal execution. In contrast, 90.0\% of the 20
initialization-activated advisory vulnerabilities are High or Critical.
Module-level code and package initialization activate 98.3\% of the
analyzed history cases. Dynamic loading is much less common, but most of its
cases perform security-sensitive actions. We also find that many fixes change
when an import becomes active instead of removing the dependency. Finally, we
derive {\bench}, 228 paired pre-fix and fixed programs covering all 11
bug types.

\end{abstract}


\maketitle

\section{Introduction}
\label{sec:introduction}

Python import does more than resolve a dependency name. When a module is
imported for the first time, Python creates the module and executes its
initialization code \cite{python-import-system}. This execution includes
module-level statements, package initializers, transitive imports,
definition-time expressions, custom loaders, and native extension
initialization. During import, code can read files, access the network, create
processes, load native libraries, or change global state before the
application calls a package API. Python import is therefore both a dependency
mechanism and an execution boundary.

A recent vulnerability in \texttt{datamodel-code-generator} illustrates this
risk. The generator accepted an attacker-controlled
\texttt{x-python-type} value and inserted it into a generated Python
annotation. A crafted value could escape the annotation and inject a statement
into a class body. The unsafe code was created during generation, but it ran
later when the generated module was imported and the class body was executed.
The fix validates the supplied value before generating the Python code
\cite{ghsa-x-python-type,ghsa-x-python-type-fix}. 
This shows that a
bug can be introduced at one point and activated later by import.

Prior work studies several related problems. Package-confusion and
software-supply-chain studies examine which dependency is selected
\cite{neupane2023packageconfusion,ohm2020backstabber}. Malicious-package
studies examine packages created or modified to perform harmful actions
\cite{guo2023maliciouspypi}. Large-scale PyPI studies examine security
problems across package collections \cite{ruohonen2021pypi}, while
vulnerability datasets connect disclosed vulnerabilities to fixing changes
\cite{ponta2019msr,bhandari2021cvefixes}. These studies cover package
selection, malicious software, broad package security, and vulnerability
fixes. We study what happens after the intended package has been selected:
what behavior import activates, why that behavior becomes a bug, and what
program information is needed to explain it.

We answer this question with {\tech}, a study of import-related bugs and
security vulnerabilities in legitimate Python software. We use two
complementary data sources. Public
security advisories provide disclosed vulnerabilities and their security
impact. Project histories capture a broader set of import-related bugs fixed
during normal development and provide the pre-fix and fixed code needed to
study their causes and repairs. For each confirmed case, we use source and patch evidence to trace execution from the import operation to the observed effect. Our study examines four aspects of these cases. We first characterize the bug landscape across both data sources. We then use the project-history corpus to study how import activates the behavior, why the bugs occur and how developers fix them, and what program information is needed to explain the pre-fix and fixed versions.

The two data sources reveal different parts of the problem. The
project-history corpus contains 1,429 confirmed bugs across 1,302
repositories. Circular imports account for 66.1\%, and the four largest bug
types together account for 93.4\%. The advisory review retains 69 relevant
cases: 31 import-related vulnerabilities and 38 application-data boundary
cases. Among the 20 vulnerabilities activated
during initialization,
18 (90.0\%) are rated High or Critical, 14 (70.0\%) allow arbitrary code
execution, and eight (40.0\%) involve generated Python. The
project-history corpus is dominated by failures that stop or disrupt
initialization, while the advisory cases show that import can also trigger
severe security effects.
The activation paths also differ. Module-level code and package
initialization activate 1,324 of the 1,347 initialization-activated
project-history bugs (98.3\%). Dynamic loading appears in only 15 cases, but 13
perform security-sensitive actions. The same bug type can also be activated
through different parts of the import process.

The fixes show that dependency structure and initialization timing both
matter. Import cycles, incorrect import targets or resolution, and
unconditional optional dependencies account for 1,236 of the 1,429 cases with
sufficient fix evidence (86.5\%). Developers often fix these problems by
changing when a dependency becomes active instead of removing it. Deferring an import is the
third most common repair and appears across all three major
dependency-related root causes.
The required program information shows a similar concentration. Most
cases need information about code that runs during import, import dependencies,
initialization order, or import resolution. The six most common requirement
combinations cover 1,240 of the 1,347 analyzed cases (92.1\%). Smaller groups
also need information about dynamic loading, native initialization, generated
code, resources, or program state.

We further derive {\bench}, a benchmark of 228 paired pre-fix and fixed
programs from confirmed cases. It covers all 11 observed bug types and
preserves the import behavior and repair that matter for each case. On
applicable pairs, Pylint reports all 30 pre-fix cycles and Pyright reports all
48 pre-fix missing imports. However, the corresponding warning disappears
after the fix in only 30.0\% and 37.5\% of the pairs, respectively. Many real
fixes change initialization order, guards, or execution timing while leaving
the structural condition in place.

This paper makes the following \textbf{\underline{contributions:}}

\begin{itemize}[leftmargin=*,nosep]

  \item We conduct a two-source empirical study of import-related bugs and
  security vulnerabilities in
  legitimate Python software. The study combines public security advisories
  with PyPI project histories and retains 69 relevant advisory cases,
  including 20 initialization-activated vulnerabilities, and 1,429
  project-history bugs across 1,302 repositories.

  \item We use the project-history corpus to characterize how these bugs are activated, why they occur, and how developers fix them. The results show that ordinary initialization
  dominates the project-history bugs, while less common activation paths can
  trigger strong security effects. Many fixes also change when a dependency is
  used instead of removing it.

  \item We identify the program information needed to explain the
  observed bugs and distinguish their pre-fix and fixed versions. A small set
  of import semantics covers most cases. Based on these findings, we derive
  {\bench}, 228 paired pre-fix and fixed programs covering all 11 observed
  bug types.

\end{itemize}

The remainder of the paper introduces Python import and related work
(Section~\ref{sec:background}), defines the study goals and research questions
(Section~\ref{sec:study}), presents the study approach
(Section~\ref{sec:approach}), reports the empirical results
(Section~\ref{sec:evaluation}), discusses their implications
(Section~\ref{sec:discussion}), and summarizes threats to validity and the
main conclusions.

\vspace{5pt}
\noindent\textbf{\underline{Open science.}}
To support replication and follow-up research, we release the code, collection
and reduction configurations, query vocabulary, confirmed-case corpus, evidence
records, classification results, {\bench}, and tool-evaluation outputs used in
this study. The repository also preserves the identifiers needed to trace each
reported case to its source advisory, repository, commit, and patch, together
with the scripts used to reproduce the reported tables and statistics. The
artifacts are available at \artifact{{\tech}}.

\section{Background and Related Work}
\label{sec:background}

This section reviews how Python import executes code and defines the security
boundary studied in {\tech}. It then summarizes related work on package
selection, malicious packages, Python package vulnerabilities, vulnerability
and fix mining, and Python program analysis.

\subsection{Python Import and Security Model}

\noindent
\textbf{\circleone{1} Import execution.}
Python import combines module search, module creation, execution, and binding
\cite{python-import-system}. For a source module, import executes the module's
top-level code in a new module namespace. Importing a submodule can first
initialize its parent package through \texttt{\_\_init\_\_.py}, and imports
performed during initialization can activate other modules. Custom loaders can
execute modules through \texttt{exec\_module}, while loading a native extension
reaches its \texttt{PyInit\_*} initialization entry point.

Python definitions can also execute code during initialization. A class
definition evaluates its header expressions, executes its body, and applies its
decorators when the class is created \cite{python-class-definitions}. Function
definitions similarly evaluate decorators and default expressions when the
definition executes \cite{python-function-definitions}. Annotation evaluation
depends on the Python version and annotation semantics and may be deferred. An
ordinary function body runs only when the function is called. These differences
determine which operations run during module initialization.

\noindent
\textbf{\circleone{2} Import attack surfaces.}
We distinguish \emph{selection} from \emph{activation}. Selection determines
which artifact satisfies an import name. Search-path manipulation, module
shadowing, typosquatting, and dependency confusion target this surface.
Activation begins after an artifact has been selected and concerns what the
artifact executes during initialization. It can occur in module-level code,
package initializers, transitive imports, definition-time evaluation, custom
loaders, or native extension initialization. Frameworks and plugin managers
provide additional activation roots when they discover and import modules
automatically.
Some behavior occurs only after initialization. For example, an application
may explicitly call a package API that later performs a dynamic import or
loads application data. Such behavior can involve an import, but execution
starts from the application operation rather than module initialization.

{\tech} focuses on behavior activated during Python, framework-driven, or
native initialization.
The behavior may occur directly in the imported module, through transitive imports,
during definition-time execution, through a loader, or during native
initialization. We treat activation as ending when further execution requires
an explicit post-import package API call. This boundary separates behavior
activated by import from functionality that is only available after import.

\noindent\textbf{\circleone{3} Security model and effects.}
Our study assumes that the intended legitimate package has already been
selected. For advisory vulnerabilities, attacker influence can come from inputs,
configuration, environment state, generated artifacts, or other state used by
the affected package. The package itself remains legitimate software.
Project-history cases also include ordinary defects that affect availability
without attacker input.
We include them as availability bugs because they stop or disrupt expected
execution, and analyze them separately from disclosed vulnerabilities.

We group observed effects into two broad classes. \emph{Authority effects} use
or change capabilities available to the process. Examples include reading or
writing files, communicating over the network, creating processes, executing
commands, accessing credentials, loading native code, or changing
security-sensitive state such as authentication, trust, or verification
settings. \emph{Availability effects} prevent normal initialization or
execution, for example through import failure, termination, circular
initialization, or excessive resource use.

We keep the concrete operation, its effect, and its possible consequence
separate. For example, an attribute assignment is an operation, disabling TLS
verification is a security-state effect, and exposure to a man-in-the-middle
attack is a possible consequence. This separation keeps program behavior and
security impact distinct.

Our main activation population contains cases in which ordinary Python,
framework-driven, or native initialization activates the relevant behavior.
We retain post-import dynamic loading, import-resolution problems, and
application-data import as separate boundary cases for the broader landscape
analysis. Application-data loading that does not use Python's module-import
machinery is not treated as import activation. Section~\ref{sec:method:confirmation}
describes how these boundaries are applied during confirmation.

The study focuses on bugs and vulnerabilities in legitimate software.
Intentionally malicious packages, typosquatting campaigns, and malware are
kept separate. We also exclude records in which the word ``import'' is
incidental and Python's module-import process does not participate in the
relevant execution.

\subsection{Related Work}

\noindent\textbf{\circleone{1} Package selection and software supply chains.}
Prior work has studied attacks that cause developers or package managers to
select an unintended package. Neupane et al.~\cite{neupane2023packageconfusion}
identify several forms of package confusion beyond simple typosquatting.
Other supply-chain studies examine how malicious packages enter dependency
trees and reach downstream software \cite{ohm2020backstabber}. These studies
mainly concern which package is selected or introduced into the dependency
chain. {\tech} starts from the intended legitimate package and studies what
happens when that package is imported and initialized.

\noindent\textbf{\circleone{2} Malicious-package analysis.}
Malicious packages have been studied across PyPI and other package ecosystems.
Ohm et al.~\cite{ohm2020backstabber} collected and analyzed malicious packages
used in real software supply-chain attacks. Guo et
al.~\cite{guo2023maliciouspypi} study the behavior and lifecycle of malicious
code in PyPI, while MalGuard~\cite{gao2025malguard} detects malicious PyPI
packages from package features and behavior.
{\tech} studies legitimate packages whose ordinary initialization
code can become security-relevant. Malicious packages are kept as a separate
population.

\noindent\textbf{\circleone{3} Python package vulnerability studies.}
Prior empirical work has studied security problems across large collections of
Python packages. Ruohonen et al.~\cite{ruohonen2021pypi}, for example, apply
security-oriented static analysis to a large PyPI snapshot and report the
distribution of detected security issues. Such studies provide a broad view of
package security.
{\tech} starts from confirmed bugs and vulnerabilities and examines their
source and fixing changes. We trace how import reaches the relevant behavior
and study its activation mechanism, root cause, repair, and required program
information.

\noindent\textbf{\circleone{4} Vulnerability and fix mining.}
Several datasets link disclosed vulnerabilities to the source changes that fix
them. Ponta et al.~\cite{ponta2019msr} manually curate links between
vulnerabilities and fixing commits. Big-Vul~\cite{fan2020bigvul} connects CVE
records with vulnerable and fixed C/C++ code, while
CVEfixes~\cite{bhandari2021cvefixes} automatically collects CVEs and their
fixing commits from open-source repositories.
{\tech} also uses fixing changes, but its project-history stream starts from
ordinary development histories rather than only known CVEs. This captures
import-related bugs that were fixed without a public vulnerability report. We
then use the pre-fix and fixed code to study why the problem occurs and how it
is repaired.

\noindent\textbf{\circleone{5} Python program analysis.}
Prior work provides program-analysis techniques and infrastructure for Python.
Monat et al.~\cite{monat2020python} develop a static analysis of Python based on
abstract interpretation. Scalpel~\cite{scalpel2022} provides reusable analyses
such as control-flow graphs, call graphs, and alias analysis for building
Python analysis tools. ModuleGuard~\cite{zhu2024moduleguard} studies namespace
conflicts among Python packages and detects cases in which different packages
provide modules with the same name.

These techniques provide useful program representations. {\tech} asks a
complementary empirical question: which program information is needed to
explain import-related bugs observed in real software? The study answers
this question by examining confirmed bugs and their fixes and measuring how
often each form of program information is needed.
Overall, prior work studies package selection, malicious packages, broad
package-security problems, vulnerability-fixing changes, and Python program
analysis. {\tech} connects these areas through import activation in legitimate
software: it traces how import reaches an observed effect and relates that
behavior to its root cause, repair, and required program information.

\section{Study Overview and Goals}
\label{sec:study}

{\tech} builds a reproducible dataset of import-related bugs from two
complementary sources: public security advisories and commit histories from
projects linked to PyPI packages. The advisory stream captures publicly
reported security vulnerabilities, while the project-history stream also
captures import-triggered failures fixed during normal software development,
such as crashes, circular imports, and resource failures.

The goal of {\tech} is to understand how Python import contributes to
these bugs and vulnerabilities. We study what problems occur, how import
activates them, what code-level conditions cause them, how developers fix them,
and what program information is needed to explain the behavior before
execution.
We organize the study around four research questions:

\begin{itemize}

  \item \textbf{RQ1: Landscape.}
  What import-related bugs and vulnerabilities appear in public
  advisories and PyPI project histories, and how are they distributed?
  We report the two discovery streams separately and characterize confirmed
  cases by scope, bug type, security effect, package or repository, time,
  severity when available, software area, and code origin. Advisory,
  fixing-commit, logical-bug, and root-cause counts remain separate because
  their relationships are not one-to-one.

  \item \textbf{RQ2: Activation.}
  How does Python import activate the relevant behavior in confirmed
  project-history bugs?
  We record what starts the import-related execution (the \emph{import
  root}) and where the relevant behavior runs (the \emph{activation
  mechanism}). Activation mechanisms include module-level code, package
  initialization, dynamic loading, native initialization, framework-driven
  import, and definition-time execution. The analysis also connects these
  mechanisms to the bug types and security effects identified in RQ1.

  \item \textbf{RQ3: Root causes and fixes.}
  What code-level root causes lead to these import-related bugs, and how do
  developers fix them?
  We compare the selected parent of each fixing commit with the fix and record
  both the underlying cause and the semantic change introduced by the repair.
  Root cause and repair are kept separate because the same cause can be fixed
  in different ways.

  \item \textbf{RQ4: Required program semantics.}
  What program information is needed to explain these bugs before
  execution?
  Using the activation, root-cause, and fix evidence, we derive the smallest
  set of program semantics needed to explain each pre-fix behavior and its
  repair. These requirements describe the program information needed to
  analyze the observed bugs.

\end{itemize}

RQ1 uses the full confirmed population from both discovery streams. 
RQ2 focuses on project-history cases with initialization-activated scope.
RQ3 further requires enough fix evidence to identify a root
cause and repair, and RQ4 requires enough activation and fix evidence to derive
the needed program semantics. Resolution, post-import, and application-data
boundary cases remain part of the RQ1 landscape.
Together, the four RQs follow a natural progression: RQ1 establishes what
import-related problems occur, RQ2 explains how import activates them, RQ3
examines why they occur and how they are fixed, and RQ4 identifies the program
information needed to analyze them.

\section{Approach}
\label{sec:approach}

\begin{figure*}[htp]
  \centering
  \includegraphics[width=\textwidth]{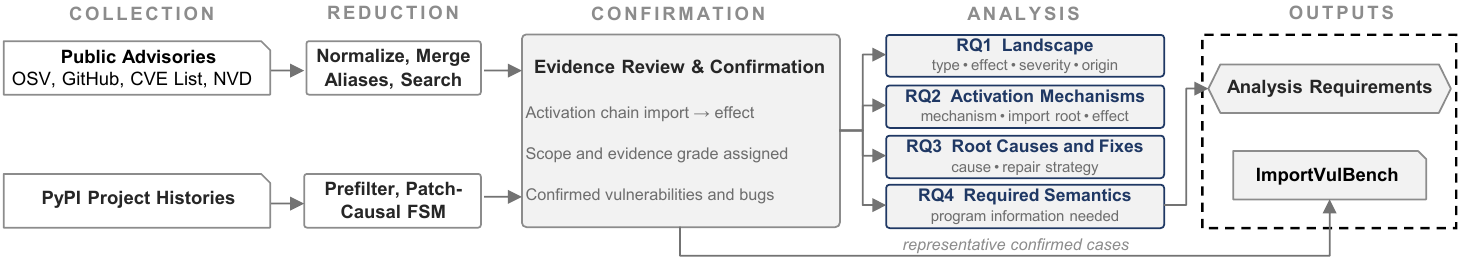}
  \caption{Overview of the {\tech} study. We collect candidate cases from
  public security advisories and PyPI project histories, reduce them through
  source-specific screening, and confirm them using source and patch evidence.
  RQ1--RQ4 characterize the landscape, activation mechanisms, root causes and
  fixes, and required program semantics. Representative confirmed cases form
  {\bench}, while RQ4 derives the analysis requirements needed to study these
  behaviors.}
  \ifdefined\Description
    \Description{The study contains five stages: collection, reduction,
    confirmation, analysis, and outputs. Public security advisories and PyPI
    project histories follow separate reduction pipelines before entering a
    common evidence-review and confirmation stage. The confirmed cases support
    four research questions covering the landscape, activation mechanisms,
    root causes and fixes, and required program semantics. Representative
    confirmed cases form ImportVulBench, while RQ4 produces the analysis
    requirements.}
  \fi
  \label{fig:study-pipeline}
\end{figure*}

Figure~\ref{fig:study-pipeline} summarizes the {\tech} workflow. We begin with
two complementary collection streams. For public security advisories, we
normalize records, merge package aliases, and search for import-related
candidates. 
For PyPI project histories, we identify candidate fixing commits
and reduce them through a metadata prefilter and a patch-causal FSM. 
The two streams then enter a common confirmation stage, where we reconstruct the
activation chain from import to effect, assign the study scope and evidence
grade, and retain confirmed vulnerabilities and bugs.

The confirmed cases support four analyses. RQ1 characterizes the overall
landscape by bug type, security effect, severity, and code origin. RQ2 examines
how import activates the relevant behavior, including the activation mechanism,
import root, and resulting effect. RQ3 identifies the code-level root causes
and the repair strategies used by developers. RQ4 determines the program
semantics needed to explain the observed behavior and distinguish the
pre-fix and fixed versions. RQ1 draws on both discovery streams, while
RQ2--RQ4 use RQ-specific subsets of the confirmed project-history cases
according to the evidence required by each analysis.
The workflow produces two outputs. Representative confirmed cases form
{\bench} for controlled evaluation. RQ4 also produces a set of analysis
requirements that summarizes the program information needed by the observed
bugs.

\subsection{Collection}
\label{sec:method:datasource}

{\tech} uses two complementary data sources: public security advisories and
commit histories of projects linked to PyPI packages. Advisories provide
publicly disclosed vulnerabilities together with their security impact.
Project histories provide a broader view of import-related bugs fixed during
normal development and include the pre-fix and fixed code needed to study
their causes and repairs. Together, the two sources cover disclosed security
vulnerabilities and bugs observed during software development.

\noindent\textbf{Security advisories.}
We collect advisories from OSV, GitHub Security Advisories, CVE List V5, and
NVD \cite{osv-data,github-global-advisories,cvelistv5,nvd-feeds}. We select
these sources based on three criteria:

\begin{itemize}
  \item \textbf{Python ecosystem coverage.}
  The source provides vulnerability records for PyPI or Python packages, so
  the collected advisories match the ecosystem studied by {\tech}.

  \item \textbf{Public and structured data.}
  The source provides public records in a structured form, allowing systematic
  collection and normalization.

  \item \textbf{Traceability to affected software.}
  The records provide package identifiers, references, or links that connect
  an advisory to the affected project, or a fixing patch. These
  links provide the evidence used during confirmation.
\end{itemize}

OSV and GitHub serve as the primary discovery sources because they provide
PyPI- and \texttt{pip}-specific records. CVE List V5 and NVD provide
additional vulnerability, severity, and weakness information for records
linked through CVE aliases. Using the 2026 snapshots, we collect 25,011 OSV
PyPI records and 5,995 reviewed GitHub \texttt{pip} advisories. CVE List V5
and NVD each contribute 196 records linked through CVE aliases, yielding
31,398 raw advisory records. Exact snapshot dates and identifiers are preserved
in the released artifact.

\noindent\textbf{PyPI project histories.}
For the project-history stream, we obtain project names from PyPI's Index API
\cite{pypi-index-api} and resolve them to source repositories through project
URL metadata \cite{pypi-core-metadata}. We use three collection rules:

\begin{itemize}
  \item \textbf{Public and accessible repository.}
  The repository must be publicly accessible because our analysis requires
  its commit history, source files, and fixing patches. This also keeps each
  confirmed case traceable to its original code.

  \item \textbf{Recent development activity.}
  The repository must have been updated within the preceding six months. This
  focuses the study on actively maintained projects and recent development
  practices.

  \item \textbf{Three-year history window.}
  For each retained repository, we collect all commits from the preceding
  three years. This provides enough recent project history to observe
  import-related fixes while limiting older code and dependency practices that
  may no longer reflect current Python packages.
\end{itemize}

We also examine the effect of the six-month activity rule.
Table~\ref{tab:activity-window} compares it with three- and twelve-month
windows. 
The three-month window gives a smaller and more recently active project set,
while the twelve-month window includes more projects with less recent
development activity. 
The six-month window balances recency and coverage.

\begin{table}[htp]
\centering
\small
\caption{Sensitivity of repository eligibility to the activity window.}
\label{tab:activity-window}
\begin{tabular}{lrr}
\hline
Activity window & Repository records & Change from six months \\
\hline
Three months & 84,082 & $-26.0\%$ \\
Six months & 113,622 & -- \\
Twelve months & 143,489 & $+26.3\%$ \\
\hline
\end{tabular}
\end{table}

The 2026 PyPI snapshot contains 878,162 project names and resolves to 501,269
unique repository records. We obtain a latest-commit date for 331,198 of these
records, of which 113,622 satisfy the six-month activity rule. Mining completes
for 110,165 of these repositories (97.0\%). The remaining 3,457 could not be
fully processed because the repositories became unavailable during collection
or their commit histories could not be retrieved completely.
For each successfully mined repository, we examine its commit history within
the preceding three-year window. Exact repository identifiers, collection
dates, and commit ranges are preserved in the released artifact. The
import-oriented retrieval applied to these histories is described in
Section~\ref{sec:method:reduction}.

For both data streams, we preserve the identifiers and references needed to
trace each retained record to its advisory, repository, commit, and patch.

\subsection{Reduction}
\label{sec:method:reduction}

The two collection streams require different reduction procedures. Advisory
records overlap across databases and can describe the same vulnerability with
different identifiers and wording. Project histories contain far more changes,
most unrelated to import behavior. We therefore reduce the two streams
separately before confirmation.

\noindent\textbf{Advisory reduction.}
We first normalize records from OSV, GitHub, CVE List V5, and NVD into a
common package-level representation. Records that refer to the same disclosed
vulnerability are grouped through explicit CVE, GHSA, PYSEC, OSV, and vendor
alias relationships. When one advisory affects several packages, we retain one
record for each package. Similar titles or package names alone are not used for
grouping.
We then retain records that match a vocabulary fixed before manual review.
The vocabulary covers import timing, Python initialization and definition
constructs, import machinery, native initialization, and generated Python. It
is preserved with the artifact. Records describing intentionally malicious packages are
separated from the legitimate-software population.

\begin{table}[htp]
\centering
\small
\caption{Advisory reduction results.}
\label{tab:advisory-reduction}
\begin{tabular}{lrr}
\hline
Step & Input & Output \\
\hline
Normalization & 31,398 & 42,378 package records \\
Alias grouping & 42,378 & 18,304 alias groups \\
Import-oriented search & 18,304 & 618 matches \\
Malware separation & 618 & 235 legitimate candidates + 383 malicious \\
\hline
\end{tabular}
\end{table}

Table~\ref{tab:advisory-reduction} summarizes the procedure.
\emph{Normalization} creates package-level records, \emph{alias grouping}
combines records connected by explicit vulnerability aliases,
\emph{import-oriented search} applies the fixed vocabulary, and
\emph{malware separation} splits legitimate-software candidates from malicious
packages.
The 235 legitimate-software candidates proceed to confirmation in
Section~\ref{sec:method:confirmation}. The 383 malicious-package records are
retained outside the main study population.

\noindent\textbf{Project-history reduction.}
Direct review of the collected project histories is impractical because most
changes are unrelated to import behavior. We first search commit messages and
changed code using an import-oriented vocabulary fixed before manual review.
This retrieval yields 1,611,273 candidate commits. We remove duplicate
repository--commit records and conservatively group commits from the same
repository when they share an issue URL or are explicit backports with the same
normalized subject. This produces 1,577,832 logical commit groups.
Following prior empirical security studies that use project evolution histories
and fixing changes to study vulnerability-related behavior
\cite{li2022multilingual,ponta2019msr}, we further reduce these logical commit
groups with a metadata prefilter followed by a patch-causal FSM.

\SetAlCapFnt{\scriptsize}
\SetAlCapNameFnt{\scriptsize}
\begin{algorithm}[htp]
\scriptsize
\setlength{\algomargin}{0.35em}
\DontPrintSemicolon
\LinesNumbered
\caption{Project-History Reduction}
\label{alg:history-reduction}

\SetKwFunction{meta}{MetadataKeep}
\SetKwFunction{patch}{PatchEvidence}
\SetKwFunction{subject}{SubjectCausalMatch}
\SetKwFunction{importev}{ImportEvidence}
\SetKwFunction{effectev}{EffectEvidence}
\SetKwFunction{repair}{RepairLink}
\SetKw{Continue}{continue}
\SetKw{Return}{return}

\KwIn{$C$: logical commit groups from eligible project histories}
\KwOut{$R$: candidates for confirmation}

$R \leftarrow \emptyset$\;

\ForEach{$c\in C$}{

    $m \leftarrow \meta(c)$\;
    \nllabel{alg:hist:meta}
    \If{$\neg m$}{
        \Continue\;
    }

    $P \leftarrow \patch(c)$\;
    \nllabel{alg:hist:patch}

    \If{$P=\emptyset$}{
        $s \leftarrow \subject(c)$\;
        \nllabel{alg:hist:fallback}
        \If{$s$}{
            $R \leftarrow R\cup\{c\}$\;
        }
        \Continue\;
    }

    $q \leftarrow q_0$\;

    \If{$\importev(P)$}{
        $q \leftarrow q_I$\;
    }
    \nllabel{alg:hist:import}

    \If{$q=q_I \land \effectev(P)$}{
        $q \leftarrow q_{IE}$\;
    }
    \nllabel{alg:hist:effect}

    \If{$q=q_{IE} \land \repair(P)$}{
        $q \leftarrow q_{IER}$\;
        $R \leftarrow R\cup\{c\}$\;
    }
    \nllabel{alg:hist:repair}
}

\Return{$R$}\;
\nllabel{alg:hist:return}

\end{algorithm}

Algorithm~\ref{alg:history-reduction} summarizes the procedure. The
metadata prefilter first removes groups without broad import-related signals
(Line~\ref{alg:hist:meta}). It then extracts patch evidence
(Line~\ref{alg:hist:patch}). When a patch is unavailable, the fallback retains
only subjects that directly connect import with the reported problem
(Line~\ref{alg:hist:fallback}). For available patches, the FSM requires import
evidence, an associated effect, and a repair that links the two
(Lines~\ref{alg:hist:import}--\ref{alg:hist:repair}). The metadata filter
removes clearly unrelated changes at low cost. The FSM avoids combining
unrelated import and failure terms from the same commit into one causal fix.
This use of fixing changes as code-level evidence follows prior manually
curated vulnerability studies \cite{ponta2019msr}.

Table~\ref{tab:history-reduction} reports the effect of the procedure. The
\emph{metadata prefilter} performs broad, low-cost screening, while the
\emph{patch-causal FSM} applies the stronger patch-level relation shown in
Algorithm~\ref{alg:history-reduction}.

\begin{table}[htp]
\centering
\small
\caption{Project-history reduction results.}
\label{tab:history-reduction}
\begin{tabular}{lrrr}
\hline
Method & Input & Output & Retained \\
\hline
Metadata prefilter & 1,577,832 & 33,437 & 2.1\% \\
Patch-causal FSM & 33,437 & 16,673 & 49.9\% \\
\hline
\end{tabular}
\end{table}

\noindent\textbf{Reduction validation.}
We audit rejected candidates to measure how often each reduction component
removes a relevant case. For each component, we draw three non-overlapping
random samples of 385 rejected logical groups and apply the same confirmation
criteria used for retained candidates. We use 385 because it is a conservative
sample size for a 95\% interval with a five-percentage-point margin at
$p=0.5$. The samples are selected without overlap using a reproducible random
procedure.
Two reviewers independently inspect the sampled records using the confirmation
criteria and then reconcile disagreements based on the saved evidence. In
Table~\ref{tab:rejection-audits}, \emph{Misses} counts sampled rejections that
meet the confirmation criteria, and \emph{Miss rate} reports this fraction with
its 95\% Wilson confidence interval \cite{wilson1927interval}.

\begin{table}[htp]
\centering
\scriptsize
\caption{Held-out rejection audits for the metadata prefilter and patch-causal FSM.}
\label{tab:rejection-audits}
\begin{tabular}{@{}llrrl@{}}
\hline
Method & Sample & Reviewed & Misses & Miss rate (95\% CI) \\
\hline
Metadata prefilter & 1 & 385 & 2 & 0.519\% (0.143--1.874\%) \\
Metadata prefilter & 2 & 385 & 0 & 0.000\% (0.000--0.988\%) \\
Metadata prefilter & 3 & 385 & 0 & 0.000\% (0.000--0.988\%) \\
Metadata prefilter & Pooled & 1,155 & 2 & 0.173\% (0.047--0.629\%) \\
\hline
Patch-causal FSM & 1 & 385 & 0 & 0.000\% (0.000--0.988\%) \\
Patch-causal FSM & 2 & 385 & 0 & 0.000\% (0.000--0.988\%) \\
Patch-causal FSM & 3 & 385 & 0 & 0.000\% (0.000--0.988\%) \\
Patch-causal FSM & Pooled & 1,155 & 0 & 0.000\% (0.000--0.331\%) \\
\hline
\end{tabular}
\end{table}

The metadata-prefilter audit finds two relevant groups among 1,155 reviewed
rejections, giving a pooled miss rate of 0.173\%. The patch-causal FSM audit
finds no relevant group among another 1,155 reviewed rejections; the 95\%
Wilson upper bound is 0.331\%.
The metadata-prefilter audit evaluates the same rule used in the main
pipeline, which retains 33,437 logical commit groups. The patch-causal FSM
then reduces these groups to 16,673 candidates for confirmation, as shown in Table~\ref{tab:history-reduction}.

\subsection{Confirmation}
\label{sec:method:confirmation}

The reduction procedures identify candidates but do not establish whether a
case satisfies the study definition. We therefore manually review all retained
advisory and project-history candidates using a common evidence protocol,
following established practice for manual analysis and curation in empirical
software-engineering studies
\cite{seaman1999qualitative,ponta2019msr}. The review establishes whether
Python import is part of the relevant execution, how execution reaches the
observed effect, and what evidence supports the case.

We divide the retained project-history candidates into three disjoint
confirmation sets, with one reviewer assigned to each set. A different
reviewer checks cases marked as uncertain or as containing more than one
independent defect. A case enters the confirmed corpus only after the saved
claim, import trigger, pre-fix behavior, repair, and exact changed locations
support one decision; the reviewers resolve remaining differences from this
evidence.

\noindent\textbf{Eligibility and activation chain.}
Reviewers inspect the advisory or fixing commit, affected source files, and
linked issues or pull requests when available. We first determine whether the
candidate describes a real defect or vulnerability and whether Python import
is material to its execution. We call the event that starts the relevant
import execution the \emph{import root}. Reviewers trace execution from this
root through package initialization, transitive imports, definition-time
execution, loaders, generated modules, or native initialization to the
observed effect.

For an initialization-activated case, we record a concrete chain of the form

\begin{center}
\small
import root $\rightarrow$ initialization construct $\rightarrow$
resolved operation $\rightarrow$ effect.
\end{center}

This chain separates behavior activated by import from behavior that becomes
reachable only after import. Cases that require a deliberate post-import
package API call, change which module is selected, or only load application
data are recorded separately according to the study boundary in
Section~\ref{sec:background}. We exclude candidates in which the import change
is unrelated to the reported problem, such as style changes, compatibility
edits, unrelated dependency updates, or documentation examples.

\noindent\textbf{Fix evidence.}
For project-history cases, we select the parent of the fixing commit that
contains the pre-fix behavior and compare it with the fixed version. We record
the changed locations and the semantic change introduced by the fix. This
comparison supports the later root-cause, repair, and required-semantics
analyses.

\noindent\textbf{Scope and evidence.}
We assign one scope from the observed execution path:
\emph{initialization-activated}, \emph{post-import}, \emph{resolution}, or
\emph{application-data boundary}. Initialization-activated cases include
behavior reached during ordinary Python initialization, framework-driven
auto-import, or native initialization. Post-import cases require an explicit
package API call before the relevant import-related behavior occurs. Resolution
cases concern which module or artifact is selected, while application-data
boundary cases load or interpret data without traversing Python's module-import
machinery.

We grade the supporting evidence as follows:

\begin{itemize}
  \item \textbf{E0:} textual evidence from an advisory, commit, issue, or
  related report;
  \item \textbf{E1:} source or patch evidence confirms the relevant
  initialization behavior;
  \item \textbf{E2:} the evidence establishes the import-to-effect chain and
  its execution boundary.
\end{itemize}

E2 is the confirmation level for initialization-activated cases.

\noindent\textbf{Confirmation results.}
Table~\ref{tab:confirmation-results} summarizes the manual review and the
scope of the retained cases. Among the 235 legitimate-software advisory
candidates, we retain 69 cases for the RQ1 landscape: 20 are activated during
initialization, seven are post-import cases, four are resolution cases, and
38 are application-data boundary cases. The first three groups form the
31 import-related advisory vulnerabilities.
For project histories, 1,429 of the 16,673 reviewed candidates are confirmed
as import-related bugs. Of these, 1,347 are activated during initialization,
one is a post-import case, and 81 are pure-resolution cases.

\begin{table}[htp]
\centering
\small
\caption{Manual-review results and scope of the retained cases.}
\label{tab:confirmation-results}
\begin{tabular}{lrrrrrr}
\hline
Source & Reviewed & Initialization-activated & Post-import & Resolution &
App.-data boundary & Retained \\
\hline
Security advisories & 235 & 20 & 7 & 4 & 38 & 69 \\
Project histories & 16,673 & 1,347 & 1 & 81 & 0 & 1,429 \\
\hline
\end{tabular}
\end{table}

The final project-history corpus is fully traceable to its supporting evidence.
All 1,429 confirmed cases have a unique identifier, repository and commit
information, the developer-reported problem, an import-to-effect explanation,
the relevant changed locations, and the corresponding stored patch.

\subsection{Analysis}
\label{sec:method:rqs}

The confirmed cases provide the evidence for the four research questions. RQ1
uses both data sources. RQ2--RQ4 use project-history cases with the evidence
needed for each analysis.

\noindent\textbf{Classification procedure.}
RQ1 bug types and RQ3 root causes do not follow a single existing taxonomy.
We reuse established CWE, advisory, and software-engineering terms when they
match the confirmed code-level pattern
\cite{cwe1000,cwe_mapping_guidance}. When existing terms do not cover a
recurring import-related pattern, we add a category based on the source, patch,
and execution evidence. We merge categories that describe the same mechanism
and split categories that cover different mechanisms. Application-data import
is treated as a scope boundary and is reported
separately in RQ1.

We refine the initial category definitions and decision rules until they are
stable enough for independent review. To obtain a conservative sample for
estimating raw agreement, we use
Cochran's formula, $n_0=z^2p(1-p)/e^2$ \cite{Cochran1977}. With 95\%
confidence ($z=1.96$), a five-percentage-point margin ($e=0.05$), and
$p=0.5$, the required sample size is 384.16. The finite-population correction
for the 1,429-case corpus gives 303 cases. We use the more conservative
uncorrected size of 385. We use non-proportional stratification: the sample
first covers every observed root-cause class and then fills the remaining
positions uniformly without replacement. Each reviewer
receives the same source and patch evidence but no final labels.

Three reviewers label the sampled cases independently before seeing one
another's decisions. After all reviews are complete, we compare the labels and
revisit each disagreement using the source, patch, and execution-chain
evidence. The reviewers resolve disagreements by consensus. If a disagreement
shows that a decision rule is unclear, we refine the rule and review affected
cases again. We then finalize the classification scheme and apply it to the
remaining cases. This process follows established practice for independent
coding and team-based review
\cite{macqueen1998codebook,macphail2016intercoder,oconner2020intercoder}.
The final bug-type and root-cause definitions are provided in
Appendices~\ref{app:classification} and~\ref{app:rootcause}.

We use the first independent labels, before reconciliation, to measure
agreement. For single-label fields, we report three-way raw agreement and
Fleiss' $\kappa$, which supports more than two reviewers
\cite{fleiss1971kappa}. For the multi-label RQ4 requirements, we report mean
pairwise Jaccard similarity \cite{jaccard1901} and per-label agreement.

\begin{table}[htp]
\centering
\small
\caption{Agreement among three independent reviewers before reconciliation.}
\label{tab:reviewer-agreement}
\begin{tabular}{lrll}
\hline
Field & $N$ & Measure & Result \\
\hline
Bug type & 385 & Raw agreement / Fleiss' $\kappa$ & 93.8\% / 0.915 \\
Import root & 385 & Raw agreement / Fleiss' $\kappa$ & 99.0\% / 0.807 \\
Activation mechanism & 385 & Raw agreement / Fleiss' $\kappa$ & 98.2\% / 0.964 \\
Root cause & 385 & Raw agreement / Fleiss' $\kappa$ & 93.8\% / 0.914 \\
Repair family & 385 & Raw agreement / Fleiss' $\kappa$ & 94.3\% / 0.861 \\
Required semantics & 349 & Mean pairwise Jaccard & 0.979 \\
\hline
\end{tabular}
\end{table}

Agreement is high across the main labels. For repairs, each reviewer first
assigns a concrete repair strategy. We then map each strategy to one of seven
repair families using a fixed mapping. The reviewers agree on the repair
family in 363 of the 385 cases (94.3\%), with Fleiss'
$\kappa=0.861$.
RQ3 reports the reconciled concrete strategy labels. The family-level measure
tests agreement on the broader kind of change; it does not claim the finer
strategy labels have the same reliability.
For RQ4, 349 of the 385 sampled cases belong to the
initialization-activated population. Three-way agreement is 100.0\% for eager
initialization, import dependency, initialization order, native initialization,
definition-time execution, generated code, and value or state information. It
is 99.4\% for dynamic loading, 98.9\% for resource information, and 96.3\%
for both import resolution and dependency and environment state.

After the agreement study, we divide the remaining confirmed cases into three disjoint
review sets. One reviewer applies the final classification scheme to each set
and records the reported problem, import trigger, pre-fix behavior, repair,
and relevant patch locations. Cases that require more evidence or need to be
split are reviewed again before entering the final corpus.

\noindent\textbf{RQ1: Landscape.}
We analyze advisories and project histories separately because they have
different discovery processes and counting units. For each confirmed case, we
record its scope, bug type, security effect, package or repository, time, and
code origin. Advisory cases also provide severity, software area, and security
consequence when available. We keep advisory records, vulnerabilities,
logical bugs, fixing commits, and root causes as separate counting units
because they are not one-to-one.
We apply the same final bug-type scheme to both data sources. Bug type
records the main code-level pattern behind the problem, while security effect
and consequence are recorded separately. Scope boundary cases, including
application-data import, are reported separately rather than forced into the
bug-type scheme. We then compare the bug-type distributions across the two
sources.

\noindent\textbf{RQ2: Activation.}
For project-history cases activated during initialization, we use the confirmed
execution chain to record both the import root and the activation mechanism.
The import root describes what starts the execution, such as an explicit
import, plugin discovery, native-dependency loading, or framework auto-import.
The activation mechanism describes where the relevant behavior runs, such as
module-level code, package initialization, dynamic loading, native
initialization, framework-driven import, or definition-time execution.
We then cross activation mechanism with bug type and security effect to study
how different problems are activated and what effects they produce.

\noindent\textbf{RQ3: Root causes and fixes.}
For cases with fix evidence, we compare the selected pre-fix parent with
the fixed version. Root cause records why the pre-fix behavior is wrong.
Repair strategy records what the patch changes. Keeping them separate allows
one root cause to map to several repairs and the same repair to appear across
different causes.

We first report the distributions of root causes and repairs and then cross the
two sets of labels to study how developers fix each type of problem. The
root-cause and repair definitions and decision rules are provided in
Appendix~\ref{app:rootcause}.

\noindent\textbf{RQ4: Required program semantics.}
RQ4 derives program requirements directly from the confirmed activation
and fix evidence rather than assigning cases to a fixed bug taxonomy. For each
case, we identify the smallest set of program semantics needed to explain the
pre-fix behavior and distinguish it from the fix. For each proposed
requirement, we ask whether the bug and fix can still be explained without that
information; if so, the requirement is removed. The 385-case reliability
sample applies the same rule independently and measures agreement on the
resulting requirement sets.
A case can require several kinds of program information. We report
both individual requirements and common combinations; the final requirement
definitions are provided in Appendix~\ref{app:requirements}.

\noindent\textbf{{\bench} construction.}
We construct {\bench} after completing the empirical analysis. For
project-history cases, we group confirmed records by bug type and retain at
most 30 cases from each category; categories with fewer than 30 cases keep all
available records. We use a reproducible selection procedure that favors cases
adding new activation mechanisms, import roots, failure patterns, scopes,
years, and repositories, with rarer failure patterns preferred before a stable
tie breaker.

For the 20 initialization-activated advisory vulnerabilities, records that
share one root cause are reduced to one representative pair when they capture
the same underlying behavior. This keeps distinct security mechanisms while
avoiding multiple pairs for advisories that describe the same defect.
Each selected case is reduced to a pre-fix/fixed program that preserves the
import entry, activation behavior, relevant program state, and semantic repair
while removing unrelated project code. Security-sensitive effects use harmless
local markers, while availability failures use deterministic exceptions or
bounded execution. Native cases preserve native initialization.
Each variant runs in a fresh process. A pair is retained only when the pre-fix
version produces its expected local marker, exception, or bounded timeout and the fixed version does not. Each pair is mapped to its source record,
and the corresponding oracle result is recorded in
\texttt{ImportVulBench/instances.jsonl}.

\section{Evaluation}
\label{sec:evaluation}

The evaluation follows the four research questions defined in
Section~\ref{sec:study}. Before presenting the results, we describe the
execution environment and the settings used for the controlled experiments.

\subsection{Experimental Setup}
\label{sec:eval:setup}

\noindent\textbf{Execution environment.}
All automated analyses and controlled experiments run on a machine with a
13th Gen Intel Core i9-13900F processor, 64~GB of memory, and Ubuntu~22.04.
Each benchmark variant runs in a fresh process so that one import does
not change the state of the next run.

\noindent\textbf{Existing-tool configuration.}
For the RQ4 comparison, we use documented built-in import checks without custom
rules or plugins. Pylint~4.0.7 uses its cyclic-import diagnostic
(\texttt{R0401}), and Pyright~1.1.404 uses
\texttt{reportMissingImports}. We report results only for cases that
match the documented purpose of each check.

\noindent\textbf{Measurements.}
RQ1--RQ3 report counts, proportions, and cross-category distributions over
their corresponding confirmed populations. RQ4 reports individual semantic
requirements and common requirement combinations. For paired pre-fix/fixed
experiments, we report the applicable and successfully analyzed cases and
whether the relevant pre-fix finding disappears after the fix.

\subsection{RQ1: Landscape}
\label{sec:eval-rq1}

RQ1 examines the import-related problems found in project histories and public
security advisories. Project histories record bugs fixed during normal
development, while advisories record publicly disclosed security
vulnerabilities. We analyze the two sources separately and then compare their
distributions.

\subsubsection{Project-History Landscape}
\label{sec:rq1:history}

Table~\ref{tab:rq1-bug-types} reports the 11 bug types found in the
two data sources and lists application-data boundary separately as an RQ1
boundary class. Percentages describe only confirmed import-related bugs and
vulnerabilities; the boundary cases do not enter either denominator.
The project-history corpus contains 1,429 confirmed bugs across 1,302
repositories. Circular imports form the largest group, with 945 cases
(66.1\%), followed by import resolution or shadowing with 177 (12.4\%),
optional-dependency handling with 114 (8.0\%), and import-time failure or
termination with 99 (6.9\%). Together, these four types account for 1,335
cases (93.4\%).
This concentration follows Python's normal import behavior. Import executes
module-level code while a module is being initialized, and recursive imports
can reach modules whose initialization is not yet complete
\cite{python-import-system}. This makes circular dependencies and failures in
initialization visible as soon as a package is imported.

\begin{table*}[htp]
\centering
\small
\caption{Bug types in the two RQ1 data sources and the separate
application-data boundary class.}
\label{tab:rq1-bug-types}
\begin{tabular}{lrrrr}
\hline
& \multicolumn{2}{c}{Project histories}
& \multicolumn{2}{c}{Advisories} \\
Category & Bugs & \% & Cases & \% \\
\hline
Generated-code execution on import & 1 & 0.1 & 8 & 25.8 \\
Import-time registration or caching & 17 & 1.2 & 3 & 9.7 \\
Security-state initialization & 0 & 0.0 & 4 & 12.9 \\
Native-extension loading & 19 & 1.3 & 1 & 3.2 \\
Dynamic module or plugin loading & 21 & 1.5 & 11 & 35.5 \\
Import resolution or shadowing & 177 & 12.4 & 4 & 12.9 \\
Import-time resource loading & 24 & 1.7 & 0 & 0.0 \\
Circular import & 945 & 66.1 & 0 & 0.0 \\
Optional-dependency handling & 114 & 8.0 & 0 & 0.0 \\
Import-time resource exhaustion & 12 & 0.8 & 0 & 0.0 \\
Import-time failure or termination & 99 & 6.9 & 0 & 0.0 \\
\hline
Confirmed import-related bugs/vulnerabilities & 1,429 & 100.0 & 31 & 100.0 \\
Application-data boundary & 0 & -- & 38 & -- \\
\hline
\end{tabular}
\end{table*}

The remaining project-history cases are more varied. Import-time registration
or caching accounts for 17 cases (1.2\%), native-extension loading for 19
(1.3\%), and dynamic module or plugin loading for 21 (1.5\%). Smaller groups
involve resources, generated code, and resource exhaustion. These mechanisms
also appear in common
Python package designs: plugins can be discovered and loaded at run time
\cite{pypa-plugins}, while optional dependencies can be absent in some
installations \cite{pypa-pyproject}.
Among the 1,429 bugs, 
1,347 are activated during Python, framework-driven, or native initialization. 
Of these, 1,314 (97.6\%) affect availability and 33 (2.4\%)
exercise authority. Thus, the project-history corpus is dominated by
bugs that stop or disrupt initialization.
The fixing commits are distributed as 73 in the 2023 portion of the
collection window, 205 in 2024, 346 in 2025, and 805 in the 2026 portion.
Because 2023 and 2026 are partial years, these counts describe only the
distribution within the collected three-year window and are not used to infer
a temporal trend.

\subsubsection{Advisory Landscape}
\label{sec:rq1:advisories}

The advisory distribution is different. The broad advisory population contains
69 relevant cases: 31 import-related vulnerabilities and 38 application-data
boundary cases. Among the 31 vulnerabilities, dynamic module or plugin loading
contributes 11 cases (35.5\%), and generated-code execution on import
contributes eight (25.8\%). The 38 application-data cases
are reported only to show the boundary encountered during advisory review.
They do not use Python's module-import machinery as the activation root and do
not enter RQ2--RQ4 or the advisory part of {\bench}. The 24 project-history
resource-loading bugs are different: they are activated during Python
initialization and remain part of the bug-type taxonomy. Resolution and
post-import cases are likewise kept in RQ1 but separated by scope.
Twenty advisory vulnerabilities have security-relevant behavior activated
during Python or native initialization. They represent 14 distinct root causes
across 10 packages in nine projects. Table~\ref{tab:rq1-advisory-summary}
summarizes their software areas, code origins, and severity.
We use the qualitative severity bands defined by CVSS v3.1
\cite{cvss31-specification}.

\begin{table*}[htp]
\centering
\small
\caption{Characteristics of the 20 advisory vulnerabilities activated during
initialization.}
\label{tab:rq1-advisory-summary}
\begin{tabular}{llrr}
\hline
Dimension & Category & Cases & \% \\
\hline
Software area
& Data and machine learning & 9 & 45.0 \\
& Serialization and code generation & 7 & 35.0 \\
& Web and networking & 2 & 10.0 \\
& Developer tooling & 1 & 5.0 \\
& Security and cryptography & 1 & 5.0 \\
\hline
Code origin
& Handwritten Python & 11 & 55.0 \\
& Generated Python & 8 & 40.0 \\
& Native extension & 1 & 5.0 \\
\hline
CVSS severity
& Critical & 9 & 45.0 \\
& High & 9 & 45.0 \\
& Medium & 2 & 10.0 \\
\hline
\end{tabular}
\end{table*}

Sixteen of the 20 vulnerabilities (80.0\%) occur in data, machine-learning,
serialization, or code-generation software. Generated Python is especially
notable: eight cases (40.0\%) occur in generated code. In these cases, input
affects Python code during generation, while the security effect appears later
when the generated module is imported. Python class bodies and related
definition-time expressions execute when the definition is evaluated
\cite{python-class-definitions}, which explains how generated definitions can
become active during import.
The advisory cases are also severe. Eighteen of the 20 initialization-activated
vulnerabilities (90.0\%) are rated High or Critical.

\subsubsection{Security Consequences}
\label{sec:rq1:consequences}

The 20 initialization-activated advisory vulnerabilities have broader
consequences than the project-history bugs. Security consequences are
multi-label. Arbitrary code execution appears in 14 cases (70.0\%),
authentication bypass in five (25.0\%), and account takeover and credential
theft in three cases each (15.0\%). Other cases involve information exposure,
session fixation, or unauthorized state changes.
These outcomes follow from code executing inside the importing Python process.
Code reached during initialization can use files, process state, credentials,
network APIs, and other resources available to that process
\cite{python-import-system}. Import-time execution can therefore change
security state or use process capabilities before normal application logic
begins.
Several advisory records describe the same underlying defect. The 20
initialization-activated advisory records reduce to 14 distinct root causes.
For example, multiple records share automatic tool-module loading, a public
default signing secret, or the same generated-code injection mechanism. We
therefore keep advisory records and root causes as separate counting units.

\subsubsection{Comparison Across the Two Sources}
\label{sec:rq1:sources}

The two sources show little overlap. None of the 1,429 project-history bugs
matches an advisory CVE in our dataset. One \texttt{skops} repository appears
in both sources, but the advisory and project-history records refer to
different bugs and different commits.

The source difference helps explain the different distributions. Project
histories capture routine fixes made during development and contain many
initialization failures. Public advisories contain disclosed security issues
and provide severity and impact information. Together, the two sources expose
common development failures and less frequent cases with stronger security
consequences.

\find{
\textbf{Finding 1.}
The two data sources reveal different parts of the import-related bug
landscape. Project histories contain 1,429 bugs across 1,302 repositories.
Circular imports account for 66.1\%, and the four largest bug types together
account for 93.4\% of the corpus. Among the 20 advisory vulnerabilities
activated during initialization, 90.0\% are High or Critical, 70.0\% allow
arbitrary code execution, and 40.0\% involve generated Python. None of the
project-history bugs overlaps with an advisory CVE in our dataset.
}

\subsection{RQ2: Activation}
\label{sec:eval-rq2}

RQ2 examines how import activates the relevant behavior. Of the 1,429
confirmed project-history bugs, we exclude 81 cases whose scope is pure import
resolution and one dynamic-loading case reached only after an explicit package
API call. 
This leaves 1,347 initialization-activated bugs.
Bug type and scope are separate fields,
so some initialization cases
can still carry an import-target or resolution bug type when the selected
target later triggers the observed effect during import.
The independent review supports these labels. Import-root agreement is 99.0\%
with Fleiss' $\kappa=0.807$, and activation-mechanism agreement is 98.2\%
with $\kappa=0.964$ (Table~\ref{tab:reviewer-agreement}).

\subsubsection{Activation Mechanisms}
\label{sec:rq2:mechanisms}

Table~\ref{tab:rq2-mechanisms} connects each RQ1 bug type with its primary
activation mechanism. Module-level code activates 865 cases (64.2\%), while
package initialization activates 459 (34.1\%). Together, these two
mechanisms account for 1,324 of the 1,347 cases (98.3\%).

\begin{table*}[htp]
\centering
\small
\caption{Primary activation mechanisms by project-history bug type. Each bug
has one primary activation mechanism. Column headings abbreviate module-level
code, package initialization, dynamic loading, native initialization,
framework-driven import, and definition-time execution.}
\label{tab:rq2-mechanisms}
\begin{tabular}{lrrrrrrr}
\hline
Bug type & Module & Package & Dynamic & Native & Framework & Definition & Total \\
\hline
Circular import & 628 & 313 & 1 & 0 & 3 & 0 & 945 \\
Generated-code execution on import & 1 & 0 & 0 & 0 & 0 & 0 & 1 \\
Import resolution or shadowing & 58 & 51 & 1 & 1 & 0 & 0 & 111 \\
Import-time failure or termination & 61 & 26 & 5 & 0 & 0 & 1 & 93 \\
Import-time resource exhaustion & 6 & 6 & 0 & 0 & 0 & 0 & 12 \\
Security-state initialization & 0 & 0 & 0 & 0 & 0 & 0 & 0 \\
Optional-dependency handling & 59 & 42 & 4 & 0 & 0 & 0 & 105 \\
Dynamic module or plugin loading & 13 & 4 & 3 & 0 & 0 & 0 & 20 \\
Import-time registration or caching & 11 & 6 & 0 & 0 & 0 & 0 & 17 \\
Import-time resource loading & 19 & 5 & 0 & 0 & 0 & 0 & 24 \\
Native-extension loading & 9 & 6 & 1 & 3 & 0 & 0 & 19 \\
\hline
Total & 865 & 459 & 15 & 4 & 3 & 1 & 1,347 \\
\hline
\end{tabular}
\end{table*}

The dominance of module-level code and package initialization follows Python's
normal import behavior. Importing a source module executes its top-level code,
while importing a regular package executes its \texttt{\_\_init\_\_.py}
\cite{python-import-system}. Circular imports and import-time failure or
termination account for 1,038 cases, and 1,028 of them (99.0\%) are activated
by module-level code or package initialization. Most observed failures
therefore come from ordinary initialization.
The smaller bug families are more varied. Of the 20 dynamic module or
plugin-loading bugs, three are activated through dynamic loading,
13 begin in module-level code, and four in package initialization. The bug
type describes the problem, while the activation mechanism describes where the
relevant behavior runs. The two are therefore separate properties.

Native-extension bugs show the same distinction. Three of the 19 cases are
activated during native initialization, while nine start from module-level code,
six from package initialization, and one from dynamic loading.
Python extension modules execute their initialization entry point when loaded,
including \texttt{PyInit\_*} for CPython extensions
\cite{python-extension-modules}. Resource-exhaustion bugs span module-level
and package activation.

\subsubsection{Import Roots}
\label{sec:rq2:roots}

The activation mechanism describes where the relevant behavior runs. The
\emph{import root}, defined in Section~\ref{sec:method:confirmation}, describes
what starts that execution. Among the 1,347 cases, 1,325 begin with an explicit
import, 15 with plugin discovery, four with native-dependency loading, and three
with framework auto-import.

The roots align closely with the less common activation mechanisms. All 15
dynamic-loading cases begin through plugin discovery, and all three
framework-driven cases begin through framework auto-import. Python packaging
supports plugin discovery through naming conventions, namespace packages, and
package metadata such as entry points \cite{pypa-plugins}. A discovered entry
can then be resolved and loaded at run time.
Root and mechanism describe different steps. An explicit import can
activate module-level code, package initialization, or native initialization.
Plugin discovery can instead lead to dynamic loading, where the target is chosen
at run time. Recording both shows how execution starts and where the relevant
behavior runs.

\subsubsection{Activation Mechanisms and Security Effects}
\label{sec:rq2:effects}

The activation mechanisms also differ in the effects they produce.
Availability failures account for 1,314 of the 1,347 cases (97.6\%), while
33 cases (2.4\%) exercise authority during import.

\begin{table}[htp]
\centering
\small
\caption{Activation mechanism by security-effect family for the 1,347
project-history bugs.}
\label{tab:rq2-mechanism-effect}
\begin{tabular}{lrrr}
\hline
Activation mechanism & Authority & Availability & Total \\
\hline
Module-level code & 14 & 851 & 865 \\
Package initialization & 6 & 453 & 459 \\
Dynamic loading & 13 & 2 & 15 \\
Native initialization & 0 & 4 & 4 \\
Framework-driven import & 0 & 3 & 3 \\
Definition-time execution & 0 & 1 & 1 \\
\hline
Total & 33 & 1,314 & 1,347 \\
\hline
\end{tabular}
\end{table}

Table~\ref{tab:rq2-mechanism-effect} shows a clear difference between ordinary
initialization and dynamic loading. Of the 865 module-level cases, 851
(98.4\%) affect availability. The same is true for 453 of the 459
package-initialization cases (98.7\%). Only 20 of these 1,324 cases (1.5\%)
exercise authority.
In our corpus, dynamic loading is strongly associated with authority
effects. Of the 15 cases activated through dynamic loading, 13
(86.7\%) exercise authority. Dynamic loading accounts for only 1.1\% of the
RQ2 population but contributes 13 of the 33 authority cases (39.4\%).

This pattern matches common plugin use. Plugin systems select and load code
that extends the host program, often based on installed packages or package
metadata \cite{pypa-plugins}. The loaded plugin runs inside the host
process, which is consistent with the high share of authority effects in this
group. Ordinary module and package initialization appears much more often in
our project histories and is mainly associated with failures that stop or
disrupt initialization.
The same mechanism can still produce different effects. Module-level code and
package initialization contain both availability and authority cases, and
availability failures also occur through dynamic loading, native
initialization, framework-driven import, and definition-time execution.
Activation mechanism and security effect therefore capture different
properties of the same execution.

\find{
\textbf{Finding 2.}
Module-level code and package initialization activate 1,324 of the 1,347
project-history bugs (98.3\%), making ordinary initialization the dominant
activation path. 
Bug type does not uniquely determine how a bug is activated.
Dynamic module or plugin loading, native-extension loading, and
import-time resource exhaustion each appear through more than one activation mechanism.
Availability failures dominate ordinary
initialization, while 13 of 15 dynamic-loading cases (86.7\%) exercise
authority and account for 39.4\% of all authority cases.
}

\subsection{RQ3: Root Causes and Fixes}
\label{sec:eval-rq3}

RQ3 examines why the confirmed import-related bugs occur and how developers
repair them. All 1,429 confirmed project-history bugs provide enough
patch evidence to identify a root cause and a repair strategy. The root-cause
and repair-strategy definitions and decision rules are provided in
Appendix~\ref{app:rootcause}.

\subsubsection{Root Causes}
\label{sec:rq3:causes}

Table~\ref{tab:rq3-root-fix} reports the 13 observed root-cause classes and the
most common repair for each class. Import cycles form the largest group, with
945 cases (66.1\%), followed by incorrect import targets or resolution with
177 cases (12.4\%). Together, these two causes account for 1,122 cases
(78.5\%). Unconditional optional dependencies add another 114 cases (8.0\%).
The three largest dependency-related causes therefore account for 1,236 of the
1,429 analyzed bugs (86.5\%).

\begin{table*}[htp]
\centering
\small
\caption{Root causes and their most common repairs for the 1,429
project-history bugs analyzed in RQ3. Corpus percentages use 1,429 as the
denominator; repair percentages use the number of cases in the corresponding
row.}
\label{tab:rq3-root-fix}
\begin{tabular}{lrrlr}
\hline
Root cause & Cases & Corpus \% & Most common repair & Within cause \% \\
\hline
Import cycle & 945 & 66.1 & Change import target & 49.1 \\
Incorrect import target or resolution & 177 & 12.4 & Change import target & 53.1 \\
Unconditional optional dependency & 114 & 8.0 & Defer import & 46.5 \\
Missing dependency or environment guard & 68 & 4.8 & Declare dependency & 33.8 \\
Eager side effect & 17 & 1.2 & Remove import-time effect & 58.8 \\
Native initialization error & 19 & 1.3 & Fix native initialization & 57.9 \\
Eager resource initialization & 24 & 1.7 & Defer initialization & 54.2 \\
Eager definition-time evaluation & 14 & 1.0 & Defer definition evaluation & 64.3 \\
Resource exhaustion & 12 & 0.8 & Defer import & 41.7 \\
Unsafe dynamic loading & 8 & 0.6 & Guard initialization & 37.5 \\
Incorrect plugin discovery & 13 & 0.9 & Change plugin discovery & 84.6 \\
Unsafe generated code & 1 & 0.1 & Change generated representation & 100.0 \\
Other project-specific cause & 17 & 1.2 & Other project-specific repair & 35.3 \\
\hline
Total & 1,429 & 100.0 & & \\
\hline
\end{tabular}
\end{table*}

The dominance of import cycles is consistent with Python's import model.
Python executes module code during import and places a module in
\texttt{sys.modules} before that execution finishes
\cite{python-import-system}. Recursive imports can therefore reach a module
that is only partly initialized. A dependency can be valid in the import graph
but still fail because the required name or state is not ready.

Incorrect targets and optional dependencies reflect two other common package
patterns. Import resolution determines which module or symbol is selected
before its code executes \cite{python-import-system}, while Python packaging
supports optional dependencies that may be absent in some environments
\cite{pypa-pyproject}. Eagerly importing such a dependency can make the whole
package fail even when only an optional feature needs it.

The smaller root-cause groups reflect other Python extension mechanisms.
Packages commonly discover plugins dynamically \cite{pypa-plugins}; class
bodies, decorators, defaults, and related expressions can execute while a
module is initialized \cite{python-class-definitions}; and native extensions
run initialization code when imported \cite{python-extension-modules}. These
cases require behavior beyond a simple import-dependency graph.

\subsubsection{Repair Strategies}
\label{sec:rq3:repairs}

Appendix~\ref{app:rootcause} defines each repair strategy and maps the
strategies to the seven families used in the agreement analysis.
Most repairs change either the import structure or when the import becomes
active. Changing the import target is the largest repair strategy, with 567
fixes (39.7\%). Breaking an import cycle accounts for 394 fixes (27.6\%), and
deferring an import accounts for 165 (11.5\%). Together, these three
strategies cover 1,126 of the 1,429 fixes (78.8\%).
Breaking a cycle and changing an import target directly change the dependency
structure or resolution target. Together they account for 961 fixes
(67.2\%). Deferring an import takes a different approach: the dependency stays
in the program but becomes active later.

This repair follows Python's execution model. Because module code runs during
import, moving an import into a function, lazy attribute lookup, or later
execution point lets the current module finish initialization before the
dependency is used \cite{python-import-system}. The failure can disappear even
when the dependency remains.
Other repairs change the conditions under which initialization proceeds.
Guarding an optional dependency appears in 47 cases (3.3\%), changing plugin
discovery in 12 (0.8\%), and guarding initialization in 51 (3.6\%). These
repairs match common package use: plugin discovery can load third-party
components dynamically \cite{pypa-plugins}, while optional dependencies can be
absent in some installations \cite{pypa-pyproject}. The repair therefore
guards, delays, or isolates the operation while keeping the feature.

\subsubsection{Root Causes and Repairs}
\label{sec:rq3:mapping}

Root cause and repair are not one-to-one. Import cycles provide the clearest
example. Among the 945 cycle cases, 464 (49.1\%) change an import
target, 394 (41.7\%) break the cycle directly, 76 (8.0\%) defer an import,
and six (0.6\%) reorder imports. The same root cause can therefore be
repaired by changing the target, changing the cycle structure, or changing
initialization timing.
Incorrect-target and resolution bugs show a similar pattern. Of the 177 cases,
94 (53.1\%) change the target, 17 (9.6\%) package the missing module or
resource, and six (3.4\%) defer the import. Some fixes therefore change when the
target is needed instead of changing the target itself.

Optional dependencies show an even stronger timing effect. Among the 114
cases, 53 (46.5\%) defer the import, 47 (41.2\%) add a guard, and five (4.4\%)
change the target. Most fixes keep the dependency but change when or under
what condition it is used.
Deferring an import cuts across all three dominant root causes. Import cycles,
incorrect targets or resolution, and unconditional optional dependencies
account for 135 of the 165 deferred-import fixes (81.8\%). Initialization
timing is therefore a common repair dimension across several bug families.

Overall, dependency structure determines which modules and packages interact,
while initialization timing determines when those dependencies are used. A
repair can change either one, which explains why the same root cause can lead
to several different fixes.

\find{
\textbf{Finding 3.}
Import cycles, incorrect import targets or resolution, and unconditional
optional dependencies account for 1,236 of the 1,429 analyzed bugs (86.5\%).
Changing an import target, breaking a cycle, and deferring an import cover 1,126
fixes (78.8\%). Deferring an import appears across all three dominant
root-cause families, showing that developers repair these bugs by changing
both dependency structure and initialization timing.
}

\subsection{RQ4: Required Program Semantics}
\label{sec:eval-rq4}

RQ4 asks what program information is needed to explain the observed
import-related bugs before execution. We use the same initialization-activated
population as RQ2. After excluding 81 pure-resolution cases and one post-import
case, 1,347 initialization-activated cases remain. We compute every requirement
count over these cases. Each analyzed case has final fix evidence. The
requirement definitions are provided in Appendix~\ref{app:requirements}.

\subsubsection{Required Program Semantics}
\label{sec:rq4:requirements}

Table~\ref{tab:rq4-semantics} reports how many of the 1,347
initialization-activated cases require each type of program information.
Requirements can overlap. All cases require knowing which code executes
automatically during import. We call this requirement \emph{eager
initialization} and treat it as the common foundation. We then examine the
additional program information needed to explain the pre-fix behavior and
distinguish it from the fix.

\begin{table}[htp]
\centering
\small
\caption{Program-information requirements for the 1,347 RQ4 cases.
Requirements can overlap, so percentages do not sum to 100\%.}
\label{tab:rq4-semantics}
\begin{tabular}{lrr}
\hline
Required program information & Cases & \% \\
\hline
Eager initialization & 1,347 & 100.0 \\
Import dependency & 1,074 & 79.7 \\
Initialization order & 947 & 70.3 \\
Dependency and environment state & 228 & 16.9 \\
Import resolution & 113 & 8.4 \\
Value or state information & 44 & 3.3 \\
Resource information & 41 & 3.0 \\
Dynamic loading & 33 & 2.4 \\
Native initialization & 28 & 2.1 \\
Definition-time execution & 14 & 1.0 \\
Generated code & 5 & 0.4 \\
\hline
\end{tabular}
\end{table}

Table~\ref{tab:rq4-semantics} shows that a small number of requirements
appear in most cases. Import dependency is needed in 1,074 cases (79.7\%),
initialization order in 947 (70.3\%), and dependency and environment state
in 228 (16.9\%). Import resolution is needed in another 113 cases (8.4\%).
The high frequency of import dependency and initialization order follows
Python's import model. Python places a module in \texttt{sys.modules} before
its code finishes executing \cite{python-import-system}. One module can
therefore observe another while the latter is only partly initialized. This
is consistent with the large \emph{Import cycle} root-cause class identified
in RQ3. Knowing the dependency alone cannot explain whether the required name
or state is ready; initialization order is also needed.

Import resolution captures another part of the import process. Python first
finds the target module and its loader and then executes the selected module
\cite{python-import-system}. Explaining cases whose behavior depends on the
selected or discovered target therefore requires knowing what an import
resolves to.
The less common requirements capture mechanisms beyond an ordinary import
dependency graph. These include run-time module discovery and loading, native
initialization, definition-time execution, generated code, external resources,
and program values or state. PyPI packages can discover and load plugins
through package metadata, namespace packages, or other discovery rules
\cite{pypa-plugins}, while class bodies and definition-time expressions can
execute during module initialization \cite{python-class-definitions}.

\subsubsection{Requirement Combinations}
\label{sec:rq4:combinations}

A single bug can require several kinds of program information. Across the
1,347 cases, we find 34 distinct minimal requirement sets. The largest combines
eager initialization, import dependency, and initialization order, covering
946 cases (70.2\%). The second combines eager initialization, dependency and
environment state, and import dependency, covering 113 cases (8.4\%). The
third combines eager initialization and import resolution, covering 86 cases
(6.4\%).

The first combination reflects the import-cycle pattern: the dependency shows
which modules interact, while initialization order shows whether one module
uses another before it is ready. The second adds whether a required package or
environment condition is present. The third covers cases where the main
question is which target an import selects.
The six most common combinations cover 1,240 cases (92.1\%). Most cases
therefore depend on a small core of program information: code that runs during
import, dependencies, initialization order, and resolution. The remaining
cases add dynamic loading, resources, native initialization, definition-time
execution, generated code, or program state.

\subsubsection{{\bench}}
\label{sec:rq4:benchmark}

Following the construction procedure in Section~\ref{sec:method:rqs}, the final
{\bench} contains 228 pre-fix/fixed pairs. Fourteen pairs represent
advisory root causes, covering 20 advisory records, and 214 pairs come from
project-history bugs.

\begin{table*}[htp]
\centering
\small
\caption{Composition of {\bench} by bug type.}
\label{tab:bench-composition}
\begin{tabular}{lrrr}
\hline
Bug type & Project histories & Advisories & Total \\
\hline
Generated-code execution on import & 1 & 7 & 8 \\
Import-time registration or caching & 17 & 3 & 20 \\
Security-state initialization & 0 & 2 & 2 \\
Native-extension loading & 19 & 1 & 20 \\
Dynamic module or plugin loading & 21 & 1 & 22 \\
Import resolution or shadowing & 30 & 0 & 30 \\
Import-time resource loading & 24 & 0 & 24 \\
Circular import & 30 & 0 & 30 \\
Optional-dependency handling & 30 & 0 & 30 \\
Import-time resource exhaustion & 12 & 0 & 12 \\
Import-time failure or termination & 30 & 0 & 30 \\
\hline
Total & 214 & 14 & 228 \\
\hline
\end{tabular}
\end{table*}

Table~\ref{tab:bench-composition} separates project-history and advisory pairs
for every bug type. Advisory pairs count distinct executable reductions rather
than the 20 advisory records they represent.
The benchmark covers all 11 observed bug types. The project-history pairs preserve common
availability and dependency-related bugs, while the advisory reductions add
security-focused patterns that are rare or absent in project histories.

\subsubsection{Existing-Tool Comparison}
\label{sec:rq4:tools}

We next examine how existing import checks relate to the program information
identified above. Table~\ref{tab:rq4-tool-capabilities} summarizes five
relevant tools. Pylint checks cyclic imports, Pyright reports unresolved
imports, deptry compares imports with declared package dependencies, and
Import Linter checks user-defined constraints over an import graph.
ModuleGuard detects namespace conflicts among modules provided by different
Python packages.

\begin{table*}[htp]
\centering
\small
\setlength{\tabcolsep}{4pt}
\caption{Existing tools related to the program information identified in RQ4.}
\label{tab:rq4-tool-capabilities}

\begin{tabular}{
p{2.5cm}
p{3.2cm}
p{4cm}
p{4cm}
}
\hline
Tool & Built-in target & Analysis focus & Related program information \\
\hline

Pylint~\cite{pylint-cyclic-import}
& Cyclic imports
& Checks module imports for cycles
& Import dependency \\

Pyright~\cite{pyright-diagnostics}
& Missing imports
& Resolves imports in the analysis environment
& Import resolution; Dependency and environment state \\

deptry~\cite{deptry-documentation}
& Dependency declarations
& Compares imports with declared dependencies
& Dependency and environment state \\

Import Linter~\cite{import-linter-contract-types}
& Import contracts
& Checks user-defined rules over the import graph
& Import dependency \\

ModuleGuard~\cite{zhu2024moduleguard}
& Module conflicts
& Detects same-name modules provided by different packages
& Import resolution; Dependency and environment state \\
\hline
\end{tabular}
\end{table*}

For the quantitative comparison, we use Pylint~4.0.7 and Pyright~1.1.404
because their built-in checks directly match conditions represented in
{\bench}. The comparison asks whether each warning tracks the pre-fix and fixed
behavior in applicable pairs. Pylint provides \texttt{R0401} for cyclic imports
\cite{pylint-cyclic-import}, while Pyright provides
\texttt{reportMissingImports} for unresolved imports
\cite{pyright-diagnostics}. We use the built-in checks without custom rules or
plugins. deptry and Import Linter cover related dependency properties, while
ModuleGuard targets cross-package module-name conflicts. These conditions do
not directly match the paired cases used in this comparison.

\begin{table}[htp]
\centering
\small
\caption{Outcomes for Pyright mappings to original package revisions.}
\label{tab:pyright-mapping-outcomes}
\begin{tabular}{lr}
\hline
Outcome & Pairs \\
\hline
Analyzed & 21 \\
Requires complete dependency environment & 14 \\
Complete pre-fix target unavailable & 9 \\
Target resolves to Python standard library & 2 \\
Source fetch timeout & 1 \\
Analysis timeout & 1 \\
\hline
Total & 48 \\
\hline
\end{tabular}
\end{table}

Applicability follows each documented diagnostic. Pylint is applied to pairs
whose confirmed condition is an import cycle. Pyright is applied to pairs with
an intentionally unresolved pre-fix import. We reassess applicability in the
original source because benchmark reduction can remove project layout and
dependency context. Table~\ref{tab:pyright-mapping-outcomes} accounts for
all 48 Pyright mappings to original package revisions. The artifact
records the commands, configurations, raw outputs, and every applicability
decision.
In the controlled cases, Pylint reports all 30 applicable pre-fix
cycles, and Pyright reports all 48 applicable missing imports. The
corresponding warning disappears after the repair in only 9 Pylint pairs
(30.0\%) and 18 Pyright pairs (37.5\%).

\begin{table*}[htp]
\centering
\small
\caption{Existing-tool results on {\bench} and original package revisions.
Only cases matching the documented check are included. Percentages use
successfully analyzed applicable pairs as the denominator. ``Paired clear''
means that the relevant warning appears before the fix and disappears after
it.}
\label{tab:rq4-existing-tools}
\begin{tabular}{lllrrrr}
\hline
Source & Tool & Built-in check & Applicable & Analyzed & Pre-fix hits & Paired clear \\
\hline
{\bench} & Pylint 4.0.7 & \texttt{R0401}/cycle
& 30 & 30 & 30/30 (100.0\%) & 9/30 (30.0\%) \\

{\bench} & Pyright 1.1.404 & \texttt{reportMissingImports}
& 48 & 48 & 48/48 (100.0\%) & 18/48 (37.5\%) \\

Original packages & Pylint 4.0.7 & \texttt{R0401}/cycle
& 30 & 30 & 14/30 (46.7\%) & 4/30 (13.3\%) \\

Original packages & Pyright 1.1.404 & \texttt{reportMissingImports}
& 23 & 21 & 12/21 (57.1\%) & 0/21 (0.0\%) \\
\hline
\end{tabular}
\end{table*}

The paired results match the repair patterns found in RQ3. Python places a
module in \texttt{sys.modules} before its initialization completes
\cite{python-import-system}. A cycle can therefore remain in the import graph
after a fix moves the relevant access or import to a later point. The cycle is
still present, but the partially initialized state is no longer used.
Likewise, a dependency can remain unavailable after a fix guards or delays its
use. In both cases, the structural condition remains while the import-time
behavior changes.

The full projects add another difference. Pylint analyzes 30 of 30 applicable
pairs and reports the relevant cycle in 14 pre-fix versions (46.7\%). The
warning disappears after four of the 30 analyzed fixes (13.3\%). Pyright
analyzes 21 of 23 applicable pairs and reports 12 pre-fix missing imports
(57.1\%); no warning disappears after the corresponding fix (0.0\%).
The controlled pairs isolate one import pattern. The original projects
also include package layout, configuration, dependencies, and other code that
can affect the check.
These results separate structural checks from import-time behavior. A
cycle or unresolved target identifies an import condition, while explaining the
bug also requires knowing how that condition is used during initialization. Other requirements found in RQ4, including dynamic loading,
resource behavior, native initialization, definition-time execution, generated
code, and program state, require information outside the two checks evaluated
here.

\subsubsection{End-to-End Case Studies}
\label{sec:rq4:cases}

The aggregate results show how often the main patterns occur. We use three
cases to show how activation, root cause, repair, and required program
information connect in real code. We select one case from the dominant
import-cycle family, one with a different activation path through plugin
discovery, and one that requires definition-time and value or state reasoning.
They illustrate common, dynamic, and specialized import behavior.


\par\smallskip
\noindent
\textbf{Case 1 (dominant case): Deferring a package export breaks an import cycle.}

\begin{tcolorbox}[
  colback=black!3,colframe=black!30,boxrule=0.3pt,arc=0.5mm,
  width=0.97\linewidth,
  left=0.8mm,right=0.8mm,top=0.5mm,bottom=0.5mm]
\scriptsize
\textbf{Package:} \texttt{palace-eval} \hfill \textbf{Role:} Dominant case\\
\textbf{Activation:} Package initialization\\
\textbf{Root cause:} Import cycle\\
\textbf{Repair:} Defer import\\
\textbf{Required information:} Eager initialization, Import dependency, Initialization order
\end{tcolorbox}

\noindent
\begin{minipage}[t]{0.48\linewidth}
\scriptsize\textbf{Before fix}
\begin{lstlisting}[style=mystyle,basicstyle=\ttfamily\scriptsize,breaklines=true,columns=fullflexible,language=Python]
from .entrypoints.palace_run import evaluate
\end{lstlisting}
\end{minipage}\hfill
\begin{minipage}[t]{0.48\linewidth}
\scriptsize\textbf{After fix}
\begin{lstlisting}[style=mystyle,basicstyle=\ttfamily\scriptsize,breaklines=true,columns=fullflexible,language=Python]
def __getattr__(name):
    if name == "evaluate":
        from .entrypoints.palace_run import evaluate
        return evaluate
    raise AttributeError(f"module 'palace' has no attribute {name!r}")
\end{lstlisting}
\end{minipage}

\noindent\textbf{\circleone{1} Activation.} Importing the package executes its initializer. The eager re-export imports the entry-point module before package initialization finishes.

\noindent\textbf{\circleone{2} Cause and repair.} Eager imports create a dependency cycle while the participating modules are still initializing. The patch moves the eager import into the later attribute-access path, so package initialization no longer creates that edge immediately.

\noindent\textbf{\circleone{3} Required information.} An analysis must model which statements run automatically during import, the dependency edge created by the eager import, and which module state is visible before initialization completes to distinguish the two versions.



\vspace{5pt}
\noindent
\textbf{Case 2 (different activation case): A broken plugin aborts discovery.}

\begin{tcolorbox}[
  colback=black!3,colframe=black!30,boxrule=0.3pt,arc=0.5mm,
  width=0.97\linewidth,
  left=0.8mm,right=0.8mm,top=0.5mm,bottom=0.5mm]
\scriptsize
\textbf{Package:} \texttt{xonsh} \hfill \textbf{Role:} Different activation case\\
\textbf{Activation:} Dynamic loading\\
\textbf{Root cause:} Incorrect plugin discovery\\
\textbf{Repair:} Change plugin discovery\\
\textbf{Required information:} Dynamic loading, Eager initialization, Import resolution
\end{tcolorbox}

\noindent
\begin{minipage}[t]{0.48\linewidth}
\scriptsize\textbf{Before fix}
\begin{lstlisting}[style=mystyle,basicstyle=\ttfamily\scriptsize,breaklines=true,columns=fullflexible,language=Python]
    for longname, _, filenames, _ in get_all_lexers():
        cls = find_lexer_class(longname)
        mod = inspect.getmodule(cls)
        val = (mod.__name__, cls.__name__)
\end{lstlisting}
\end{minipage}\hfill
\begin{minipage}[t]{0.48\linewidth}
\scriptsize\textbf{After fix}
\begin{lstlisting}[style=mystyle,basicstyle=\ttfamily\scriptsize,breaklines=true,columns=fullflexible,language=Python]
    for longname, _, filenames, _ in _safe_iter(get_all_lexers()):
        try:
            cls = find_lexer_class(longname)
            mod = inspect.getmodule(cls)
            val = (mod.__name__, cls.__name__)
        except Exception:
            continue
\end{lstlisting}
\end{minipage}

\noindent\textbf{\circleone{1} Activation.} Import-time plugin discovery enumerates and loads Pygments entry points. A broken third-party entry point can therefore stop discovery.

\noindent\textbf{\circleone{2} Cause and repair.} Plugin discovery selects, loads, or orders plugins incorrectly during initialization. The patch catches exceptions raised while plugin entries are enumerated or resolved, so one broken entry no longer stops the discovery pass.

\noindent\textbf{\circleone{3} Required information.} An analysis must model the modules reached through run-time plugin discovery, which statements run automatically during import, and which module or entry point the loader resolves to distinguish the two versions.



\vspace{5pt}
\noindent
\textbf{Case 3 (specialized case): An API key is captured at definition time.}

\begin{tcolorbox}[
  colback=black!3,colframe=black!30,boxrule=0.3pt,arc=0.5mm,
  width=0.97\linewidth,
  left=0.8mm,right=0.8mm,top=0.5mm,bottom=0.5mm]
\scriptsize
\textbf{Package:} \texttt{scitex-genai} \hfill \textbf{Role:} Specialized case\\
\textbf{Activation:} Definition-time execution\\
\textbf{Root cause:} Eager definition-time evaluation\\
\textbf{Repair:} Defer definition evaluation\\
\textbf{Required information:} Eager initialization, Definition-time execution,
Value or state information
\end{tcolorbox}

\noindent
\begin{minipage}[t]{0.48\linewidth}
\scriptsize\textbf{Before fix}
\begin{lstlisting}[style=mystyle,basicstyle=\ttfamily\scriptsize,breaklines=true,columns=fullflexible,language=Python]
def __init__(
    self,
    api_key: Optional[str] = os.getenv("GROQ_API_KEY"),
):
    ...
\end{lstlisting}
\end{minipage}\hfill
\begin{minipage}[t]{0.48\linewidth}
\scriptsize\textbf{After fix}
\begin{lstlisting}[style=mystyle,basicstyle=\ttfamily\scriptsize,breaklines=true,columns=fullflexible,language=Python]
def __init__(
    self,
    api_key: Optional[str] = None,
):
    api_key = api_key or os.getenv("GROQ_API_KEY")
\end{lstlisting}
\end{minipage}

\noindent\textbf{\circleone{1} Activation.} Importing \texttt{\_Groq.py} creates the
constructor and evaluates its default argument. The environment value is
therefore captured before any constructor call.

\noindent\textbf{\circleone{2} Cause and repair.} The default argument reads the API key too
early, so later environment updates are ignored. The patch changes the default
to \texttt{None} and reads the environment inside the constructor.

\noindent\textbf{\circleone{3} Required information.} An analysis must model code that runs
during import, default-argument evaluation, and the environment value visible
before and after the constructor call.

\find{
\textbf{Finding 4.}
Beyond the shared need to model eager initialization, import dependency is
required in 79.7\% of the analyzed cases and initialization order in 70.3\%.
The six most common requirement combinations cover 92.1\% of the cases.
Existing cycle and missing-import checks capture import structure, but the
same warning often remains after the bug is fixed because the repair changes
initialization order, guarding, or execution timing.
}

\section{Discussion}
\label{sec:discussion}

The results of {\tech} show that Python import is not only a dependency
mechanism, but also an execution boundary that shapes reliability and security.
Across the four research questions, we observe clear patterns in how
import-related bugs are activated, why they occur, how developers repair them,
and what program information is needed to explain them. These findings have
direct implications for Python package development, security review, and
future import analysis.

\noindent\textbf{Import-time behavior is also a security concern.}
Project histories contain 1,429 confirmed bugs across 1,302 repositories and
are dominated by availability failures. In contrast, 18 of the 20 advisory
vulnerabilities activated during initialization (90.0\%) are rated High or
Critical, and 14 (70.0\%) allow arbitrary code execution. Other cases cause
authentication bypass, credential exposure, or unsafe security state.
Import-time behavior is therefore more than a source of startup
failures. It can also perform security-sensitive actions before normal
application logic begins. Security review should include module and package
initialization.

\noindent\textbf{Common and security-sensitive activation paths differ.}
Module-level code and package initialization activate 1,324 of the 1,347 RQ2
cases (98.3\%). Dynamic loading appears in only 15 cases (1.1\%), but 13 of
them (86.7\%) exercise authority and account for 39.4\% of all 33 authority
cases.
These results suggest two analysis priorities. Module and package
initialization provide the broadest coverage. 
Dynamic loading also deserves focused security analysis because it
accounts for a large share of authority effects despite its small number.

\noindent\textbf{Bug type and activation mechanism should be kept separate.}
The same bug type can be activated in different ways. Among the 20 dynamic
module or plugin-loading bugs, three are activated through dynamic loading, 13
through module-level code, and four through package initialization.
Native-extension and resource-exhaustion bugs also span multiple activation
mechanisms. The same activation mechanism can also lead to different effects.
Module-level code and package initialization produce both availability and
authority effects.
Import-related bugs should record what goes wrong, how import
activates the behavior, and what effect follows. These are different parts of
the same execution.

\noindent\textbf{Initialization timing matters alongside dependency structure.}
Import cycles, incorrect import targets or resolution, and unconditional
optional dependencies account for 1,236 of the 1,429 RQ3 cases (86.5\%).
Deferring an import is the third most common repair, with 165 fixes (11.5\%).
Of these, 135 (81.8\%) occur in the three largest dependency-related root-cause
families. In many of these cases, the dependency remains but becomes active
later.
Import analysis therefore needs both dependency structure and timing. It
must know which modules depend on one another and when those dependencies are
used during initialization. This matches RQ4, where initialization order is
required in 947 cases (70.3\%).

\noindent\textbf{Existing import checks capture structure, but not always the
fixed behavior.}
On {\bench}, Pylint reports all 30 applicable pre-fix cycles, and Pyright
reports all 48 applicable missing imports. After the fixes, the corresponding
warning disappears in only 9 Pylint pairs (30.0\%) and 18 Pyright pairs
(37.5\%). The same pattern appears in the original projects: the Pylint warning
disappears after four of 30 analyzed fixes, while none of the 21 analyzed
Pyright pairs clears after the fix.
These checks report the structural conditions they target. Many fixes
instead change initialization order, guards, or execution timing. A structural
warning can remain after the import-time bug is repaired.

\noindent\textbf{Most cases depend on a small set of program information.}
Import dependencies are needed in 1,074 RQ4 cases (79.7\%), initialization
order in 947 (70.3\%), and dependency and environment state in 228 (16.9\%).
The six most common requirement combinations cover 1,240 of the 1,347 analyzed
cases (92.1\%).
Code that runs during import, import dependencies, initialization order,
and resolution should therefore receive the highest priority in import
analysis. Smaller groups also require environment state, dynamic loading,
resources, native initialization, definition-time execution, generated code,
or program state.

\noindent\textbf{Generated Python is part of the executable package surface.}
Generated Python appears in eight of the 20 advisory vulnerabilities activated
during initialization (40.0\%). In these cases, untrusted input first affects
generated Python code, and the security effect appears later when the generated
module is imported.
Security analysis of code-generating packages should therefore examine
both the generator and the generated module. The source of unsafe code and the
point where it executes can be different.

\noindent\textbf{The fixes suggest simple ways to reduce import-time risk.}
Developers often break import cycles, defer imports, or change import targets.
Optional dependencies are guarded or delayed. Plugin-discovery failures are
isolated. Security-sensitive values are moved out of module-level
initialization.
These fixes suggest a simple design rule: keep import-time work small and
predictable. Delay work that depends on optional dependencies, dynamically
selected modules, runtime resources, or security-sensitive state until that
information is needed.

\section{Threats to Validity}
\label{sec:threats}

\noindent\textbf{Construct validity.}
The main construct threat is deciding what counts as an import-related bug.
The word ``import'' can refer to Python module loading or to unrelated actions
such as loading application data. A commit can also contain import-related and
failure-related changes that are not causally connected.
We address this by confirming cases from source and patch evidence rather than
keywords alone. For each confirmed case, we reconstruct the execution chain
from the import root to the observed effect. We keep bug type,
activation mechanism, security effect, root cause, repair, and required
semantics as separate fields. Application-data import is treated as a boundary
class rather than a Python import-activation bug type; it is reported in RQ1
but excluded from RQ2--RQ4. Resolution and post-import cases are also recorded
separately by scope.

\noindent\textbf{Internal validity.}
A fixing commit can contain several changes, and the selected parent may not
exactly match the affected release. For each project-history case, we compare
the selected pre-fix parent with the fixed version and inspect the relevant
source, patch, issue, pull request, or advisory when available. Cases without
enough evidence remain outside the analysis that requires that evidence.
Automatic reduction can also remove relevant candidates before confirmation.
We therefore audit rejected candidates from both the metadata prefilter and the
patch-causal FSM using independent random samples. The observed miss rates and
95\% Wilson confidence intervals are reported in Section~\ref{sec:method:reduction}. 
These audits
evaluate the same reduction rules used in the main project-history pipeline.
The rejection audits begin with the post-search candidate pool, so they do not
measure relevant commits missed by the initial retrieval vocabulary.

\noindent\textbf{External validity.}
The collected data do not cover every Python import-related bug or
vulnerability. The advisory stream contains publicly reported vulnerabilities,
while the project-history stream includes PyPI projects with public,
accessible, recently active repositories. Repository histories are limited to
a three-year window.
The findings apply most directly to disclosed vulnerabilities
and actively maintained PyPI projects under these collection criteria. Other
package ecosystems, older projects, private repositories, and undisclosed
vulnerabilities may show different patterns.

\noindent\textbf{Reliability.}
Manual confirmation and classification require human judgment.
We measure reviewer agreement on 385 confirmed cases. Three reviewers
label these cases independently before reconciliation and resolve disagreements
by returning to the source, patch, and execution-chain evidence. Agreement is
measured with three-way raw agreement and Fleiss' $\kappa$ for single-label
fields and pairwise Jaccard similarity for RQ4 requirement sets; the results
are reported in Table~\ref{tab:reviewer-agreement}. The final classification
rules and decision boundaries are also provided in the appendix.
The agreement study measures classification consistency after confirmation; it
does not provide an independent confirmation decision for every retained
candidate. Outside the agreement sample, the remaining confirmed cases are
divided into three disjoint review shards, and the assigned reviewer applies
the final classification scheme
while recording the developer-reported problem, import trigger, repair evidence, and exact
patch locations. Cases without enough evidence for one decision are excluded
from the confirmed corpus rather than assigned a forced label.
We preserve the source records, repository and commit identifiers, patches,
final labels, and data used to generate the reported tables and figures.

\noindent\textbf{Artifact availability.}
The code, fixed configurations, query vocabulary, confirmed-case corpus,
evidence patches, benchmark, tool outputs, and validators are available at
\url{https://github.com/awen-li/ImportMine}. The formal corpus is under
\texttt{ImportVulCorpus}, and the benchmark is under \texttt{ImportVulBench}.
These two directories are the source of truth for the reported corpus and
benchmark results.

\section{Conclusion}
\label{sec:conclusion}

This work presents {\tech}, a study of import-related bugs and security
vulnerabilities. We combine public security advisories with project
histories from PyPI packages to study disclosed vulnerabilities and bugs fixed
during normal development. Using source, patch, and execution evidence, we
confirm import-to-effect chains and analyze activation mechanisms, root causes,
fixes, and required program semantics.

Most project-history bugs arise from ordinary module and package
initialization, while less common mechanisms such as dynamic loading account
for many security-sensitive cases. Import cycles, incorrect import targets,
and optional dependencies dominate the observed root causes, and many fixes
change when a dependency becomes active instead of removing it. A small set of program information explains most confirmed cases, especially
code that runs during import, import dependencies, initialization order, and
resolution.
These findings show that Python import is an execution boundary as well as a
dependency mechanism. The study also provides {\bench}, a set of paired
pre-fix and fixed programs derived from confirmed cases, and a set of
analysis requirements grounded in the observed bugs and fixes.
These
artifacts support future work on detecting and understanding import-related
bugs and security vulnerabilities.

\bibliographystyle{ACM-Reference-Format}
\bibliography{reference}

\newpage
\appendix
\section*{Appendix} \label{sec:app}

\section{Bug-Type Classification and Decision Rules}
\label{app:classification}

This appendix defines the final bug-type classification used in RQ1.
We begin with established CWE and advisory terms when they match the confirmed
code-level pattern \cite{cwe1000,cwe_mapping_guidance}. When those terms do
not cover a recurring import-related mechanism, the reviewers add a category
from the source, patch, and execution evidence. Candidate categories are
merged when they describe the same mechanism and split when one category hides
distinct mechanisms. The names reuse established concepts such as circular
imports, resolution, optional dependencies, and resource exhaustion, but all
11 decision boundaries are defined for this study's import-execution setting.
Three reviewers apply the initial scheme to the
independent 385-case sample described in Section~\ref{sec:method:rqs}. After
agreement reconciliation and rule refinement, we finalize the scheme before
coding the remaining cases. Bug type records the code-level pattern, while
security effect, consequence, and scope are recorded separately.

\subsection{General Decision Rules}

Reviewers follow the same rules for both data sources:

\begin{itemize}
  \item \textbf{Use the observed mechanism.}
  The label follows the confirmed source and execution evidence. Advisory
  wording, package name, severity, and data source do not determine the type.

  \item \textbf{Prefer the most specific supported type.}
  When several patterns appear in one case, we use the most specific
  code-level mechanism supported by the evidence. A general label such as
  import-time failure is used only when no more specific type explains the
  problem.

  \item \textbf{Separate mechanism from effect.}
  Bug type records what goes wrong in the import-related behavior. Effects such
  as availability failure or authority exercise and consequences such as code
  execution or authentication bypass are recorded separately.

  \item \textbf{Separate resolution from activation.}
  A case is classified as import resolution or shadowing when the defect
  changes which module or artifact is selected. Code that runs after selection
  is classified by the mechanism that causes the observed problem.

  \item \textbf{Use one primary bug type.}
  Each confirmed case receives one primary bug type. Other properties of the
  case are captured by the activation, effect, consequence, root-cause, and
  repair labels.
\end{itemize}

\subsection{Bug-Type Definitions}

Table~\ref{tab:bugtype-codebook} gives the final categories and the rule used
to assign each one.

\begin{table*}[htp]
\centering
\small
\caption{Bug-type classification used for RQ1.}
\label{tab:bugtype-codebook}
\begin{tabular}{p{0.27\textwidth}p{0.65\textwidth}}
\hline
Bug type & Decision rule \\
\hline

Generated-code execution on import
&
Generated Python contains behavior that becomes active when the generated
module is imported. The generation step and the later import are recorded
separately. \\

Import-time registration or caching
&
Initialization registers, caches, or stores shared state, and that
initialization behavior is the main cause of the problem. \\

Security-state initialization
&
Initialization creates or changes security-sensitive state, such as trust,
authentication, authorization, signing, or verification state. \\

Native-extension loading
&
The relevant problem occurs while a native extension or native dependency is
loaded or initialized, including behavior reached through
\texttt{PyInit\_*}. \\

Dynamic module or plugin loading
&
The problem is caused by run-time module discovery, plugin discovery, dynamic
module selection, or loading of the selected module. \\

Import resolution or shadowing
&
The defect involves an incorrect import target, symbol, path, or resolution
decision. Pure artifact-selection cases are marked with Resolution
scope and excluded from RQ2; cases whose selected target then triggers an
initialization effect can remain in the activation analysis. \\

Import-time resource loading
&
The relevant behavior loads or interprets an external resource during
Python or native initialization, and that operation causes the observed
problem. \\

Circular import
&
The problem is caused by a cyclic import dependency while the participating
modules or packages are still being initialized. \\

Optional-dependency handling
&
The code assumes that an optional dependency is present or handles its absence
incorrectly during import. \\

Import-time resource exhaustion
&
Initialization performs excessive or unbounded work, memory use, recursion,
I/O, or other resource consumption. \\

Import-time failure or termination
&
Initialization raises an exception, terminates, or otherwise prevents normal
import, and no more specific category above explains the failure. \\

\hline
\end{tabular}
\end{table*}

Application-data import is not a bug type in this classification. It
is a scope boundary used for advisory records in which an application feature
loads or interprets data without traversing Python's module-import machinery.
These cases are reported separately in RQ1 and are not included in the
activation analyses.

\section{Root-Cause and Repair Classification}
\label{app:rootcause}

This appendix provides the decision rules and category boundaries used for
the RQ3 root causes and repair strategies.
We start from established software and security terms when they match the
confirmed cause. When repeated cases do not fit those terms, the reviewers add
a category from the source, patch, and pre-fix/fixed evidence. Candidate
categories are merged or split until each has a clear definition and boundary.
The names reuse established concepts where they fit, but the 13 root-cause
boundaries and their precedence are defined for this study.
After agreement reconciliation and rule refinement, we finalize the scheme
before coding the remaining cases. Reviewer agreement is measured on the
independent 385-case sample described in Section~\ref{sec:method:rqs}.

\subsection{General Decision Rules}

Reviewers follow four rules when assigning a root cause:

\begin{itemize}
  \item \textbf{Use the pre-fix condition.}
  The root cause describes what is wrong in the vulnerable or buggy version,
  not the symptom that appears later.

  \item \textbf{Use the fix as supporting evidence.}
  The patch helps identify which condition the developer changed to remove the
  problem.

  \item \textbf{Prefer the most specific supported cause.}
  A specific cause such as an import cycle or incorrect plugin discovery is
  preferred over a general category such as an eager side effect.

  \item \textbf{Keep root cause and repair separate.}
The root cause records why the pre-fix behavior is wrong, while the repair
records what the patch changes. The same root cause can therefore map to
different repairs.

\end{itemize}

\subsection{Root-Cause Definitions}

Table~\ref{tab:rootcause-codebook} gives the final root-cause categories used
for RQ3.

\begin{table*}[htp]
\centering
\small
\caption{Root-cause classification used in RQ3.}
\label{tab:rootcause-codebook}
\begin{tabular}{p{0.28\textwidth}p{0.64\textwidth}}
\hline
Root cause & Decision rule \\
\hline

Import cycle
&
Two or more modules or packages form a cyclic import dependency, and the
problem occurs because one of them is used before initialization completes. \\

Incorrect import target or resolution
&
The import resolves to the wrong module, package, symbol, path, or resource,
or the code refers to an incorrect import target. \\

Unconditional optional dependency
&
An optional dependency is imported or required without first handling the case
in which it is unavailable. \\

Unsafe dynamic loading
&
Run-time loading selects or executes a module in a way that causes the observed
problem. \\

Incorrect plugin discovery
&
Plugin enumeration, selection, loading, or failure handling is incorrect and
causes the observed behavior. \\

Eager resource initialization
&
A file, model, credential, device, or other resource is initialized during
import when that work should occur later or under different conditions. \\

Resource exhaustion
&
Import performs unbounded or excessive computation, recursion, memory use,
I/O, or other resource-consuming work. \\

Native initialization error
&
The problem is caused by incorrect behavior while a native extension or native
dependency is initialized. \\

Eager definition-time evaluation
&
A class or function definition evaluates an expression too early, such as a
base, decorator, default value, or related definition-time expression. \\

Eager side effect
&
Module initialization performs a side effect too early, and no more specific
root-cause category above explains the behavior. \\

Unsafe generated code
&
Generated Python contains an unsafe construct or value that becomes relevant
when the generated module is imported. \\

Missing dependency or environment guard
&
Import-time behavior assumes a required dependency, configuration value, or
environment condition without checking that it is available or valid. \\

Other project-specific cause
&
The evidence supports a clear project-specific root cause that does not fit any
of the recurring categories above. This category is used only after the other
categories have been ruled out. \\

\hline
\end{tabular}
\end{table*}

\subsection{Repair Strategy Definitions}

Table~\ref{tab:repair-strategy-codebook} defines the concrete repair
strategies and their fixed mapping to the seven repair families used for
reviewer agreement. Strategies describe the operation performed by the patch,
not the pre-fix cause. For example, changing a target selects a different
module or symbol, breaking a cycle removes or restructures a cyclic dependency,
and deferring an import keeps the dependency but activates it later.

\begin{table*}[htp]
\centering
\scriptsize
\caption{Repair strategies and their repair-family mapping.}
\label{tab:repair-strategy-codebook}
\begin{tabularx}{\textwidth}{p{0.22\textwidth}Xp{0.24\textwidth}}
\hline
Repair strategy & Definition & Repair family \\
\hline
Break import cycle
& Change an import edge or code ownership so cyclic initialization no longer exposes partial state.
& Import structure and timing \\
Defer import
& Move the import to the later function, callback, or lazy path that needs it.
& Import structure and timing \\
Change import target
& Change the imported module, symbol, path, or package layout used by resolution.
& Import structure and timing \\
Reorder imports
& Change import order so the required state is initialized first.
& Import structure and timing \\
Remove import
& Remove the eager import from the initialization path.
& Import structure and timing \\
Defer initialization
& Postpone resource or object initialization until explicit use.
& Import structure and timing \\
Defer definition evaluation
& Postpone a default, annotation, decorator, or other definition-time expression.
& Import structure and timing \\
Guard optional dependency
& Check or catch the absence of an optional dependency before using it.
& Availability and failure handling \\
Guard initialization
& Add a condition or failure handler around initialization.
& Availability and failure handling \\
Add fallback
& Use a safe alternative when the preferred initialization path is unavailable.
& Availability and failure handling \\
Package required artifact
& Include the module or resource required by the installed package.
& Dependency and resource provisioning \\
Declare dependency
& Declare a required distribution so supported installations provide it.
& Dependency and resource provisioning \\
Fix native initialization
& Correct or guard native-library loading or initialization.
& Dependency and resource provisioning \\
Change plugin discovery
& Change how plugins are enumerated, selected, or loaded.
& Dynamic loading \\
Restrict dynamic loading
& Limit which dynamically selected module or plugin may be loaded.
& Dynamic loading \\
Bound resource use
& Bound or serialize work that could exhaust resources during initialization.
& Effect and state control \\
Remove import-time effect
& Remove an externally visible effect from module initialization.
& Effect and state control \\
Remove global mutation
& Remove a module-level mutation of shared state.
& Effect and state control \\
Change generated representation
& Change how generated initialization code represents a value.
& Generated-code repair \\
Other project-specific repair
& Apply a supported repair that does not fit a recurring strategy above.
& Other \\
\hline
\end{tabularx}
\end{table*}

\section{Required Program Semantics Definitions}
\label{app:requirements}

This appendix defines the program semantics used in RQ4. Each
requirement captures information needed to explain the pre-fix behavior and
distinguish it from the corresponding fix. A case can require more than one
type of information. We keep the smallest set supported by the activation and
fix evidence. Table~\ref{tab:rq4-requirements} gives the final definitions.

\begin{table*}[htp]
\centering
\small
\caption{Program-information requirements used in RQ4.}
\label{tab:rq4-requirements}
\begin{tabular}{p{0.27\textwidth}p{0.65\textwidth}}
\hline
Requirement & Definition \\
\hline

Eager initialization
&
Which code executes automatically while a module or package is
initialized. \\

Import dependency
&
Which modules or packages depend on one another through imports. \\

Initialization order
&
The order in which modules or packages are initialized and the state
visible at each point. \\

Import resolution
&
Which module, package, symbol, or artifact an import resolves to. \\

Dependency and environment state
&
Whether required dependencies, configuration, or environment
conditions are present or valid. \\

Value or state information
&
Program values or state that affect import-time behavior or its security
effect. \\

Dynamic loading
&
Modules or targets selected through run-time discovery or loading. \\

Resource information
&
Files, models, devices, credentials, or other resources used during
initialization. \\

Native initialization
&
Behavior that occurs while native code or an extension module is loaded
or initialized. \\

Definition-time execution
&
Code executed while creating classes or functions, such as bases,
decorators, defaults, or other definition-time expressions. \\

Generated code
&
The link between generated Python code and the behavior that occurs when
the generated module is later imported. \\

\hline
\end{tabular}
\end{table*}

For each case, we keep a requirement only when it is needed to
explain the pre-fix/fixed difference. For example, if delaying one dependency
fixes a circular import, the dependency alone does not explain the change;
initialization order is also required.

\end{document}